\documentclass{aa}
\usepackage{graphicx}
\usepackage{txfonts}
\usepackage[utf8]{inputenc}
\usepackage{xcolor}
\usepackage{amssymb}
\usepackage{graphicx}
\graphicspath{{figures/}}
\usepackage[colorlinks=true,linkcolor=blue,citecolor=blue, urlcolor=teal]{hyperref}
\usepackage{textcomp}
\usepackage{booktabs}
\usepackage{bm}
\usepackage{physics}
\usepackage{dsfont}
\usepackage{gensymb}
\usepackage{mathtools}
\usepackage{float}
\usepackage{algorithm} 
\usepackage{algpseudocode}
\usepackage{dblfloatfix}
\usepackage{natbib}
\usepackage{nccmath}
\bibpunct{(}{)}{;}{a}{}{,}

\newcommand{\rom}[1]{%
  \textup{\uppercase\expandafter{\romannumeral#1}}%
}

\usepackage{ulem}

\newcommand\ab{\bm{a}}
\newcommand\mb{\bm{m}}
\newcommand\rb{\bm{r}}
\newcommand\Db{\bm{D}}
\newcommand\mub{\bm{\mu}}
\newcommand\sigmab{\bm{\sigma}}
\newcommand\thetab{\bm{\theta}}

\title{On the multiphase structure of the neutral interstellar medium -
	I. Phase separation using ROHSA}

\begin{document}
      \title{ROHSA: Regularized Optimization for Hyper-Spectral Analysis
      \thanks{{\tt ROHSA} is available in free access via the following web page : \url{https://github.com/antoinemarchal/ROHSA}. }}
        \subtitle{Application to phase separation of 21 cm data}
   \author{Antoine Marchal\inst{1,2}
          \and
          Marc-Antoine Miville-Deschênes \inst{1}
          \and
          François Orieux \inst{2,3}
	      \and
          Nicolas Gac \inst{3}
          \and
          Charles Soussen \inst{3}
	      \and
          Marie-Jeanne Lesot \inst{4}
          \and
          Adrien Revault d'Allonnes \inst{4}
          \and
          Quentin Salomé \inst{5}
          }
   \institute{
   AIM, CEA, CNRS, Université Paris-Saclay, Université Paris Diderot, Sorbonne Paris Cité, F-91191 Gif-sur-Yvette, France.
   \and
   Institut d’Astrophysique Spatiale, CNRS UMR 8617,
   Université Paris-Sud 11, Batiment 121, 91405, Orsay, France. \\
   \email{antoine.marchal@ias.u-psud.fr}
   \and
   Laboratoire des Signaux et Systèmes (CNRS, CentraleSupélec, Univ. Paris-Sud),
   Univ. Paris-Saclay, 91192, Gif-sur-Yvette, France.
   \and
   LIP6, Université Pierre et Marie Curie-Paris 6, UMR7606, 4 Place Jussieu Paris
   cedex 05, 75252, France.
   \and
   Instituto de Radioastronom\'ia y Astrof\'isica, Universidad Nacional Aut\'onoma
   de M\'exico, 58089 Morelia, Mexico.
   }
   \date{Received February 21, 2019; -- Accepted April 26, 2019}

  \abstract
{Extracting the multiphase structure of the neutral interstellar medium (ISM)
is key to understand the star formation in galaxies. The radiative
condensation of the diffuse warm neutral medium producing a thermally unstable
lukewarm medium and a dense cold medium is closely related to the initial step which leads the atomic-to-molecular (HI-to-H$_2$) transition and the formation of molecular clouds. Up to now the mapping of these phases out of 21\,cm emission hyper-spectral cubes has remained elusive mostly due to the velocity blending of individual cold structures present on a given line of sight. As a result, most of the current knowledge about the HI phases rests on a small number of absorption measurements on lines of sight crossing radio sources.}
{The goal of this work was to develop a new algorithm to perform separation of diffuse sources in hyper-spectral data. Specifically the algorithm was designed in order to address the velocity blending problem by taking advantage of the spatial coherence of the individual sources. The main scientific driver of this effort was to extract the multiphase structure of the HI from 21\,cm line emission only, providing a mean to map each phase separately, but the algorithm developed here should be generic enough to extract diffuse structures in any hyper-spectral cube.}
{We developed a new Gaussian decomposition algorithm named {\tt ROHSA} (Regularized Optimization for Hyper-Spectral Analysis) based on a multi-resolution process from coarse to fine grid. {\tt ROHSA} uses a regularized non-linear least-square criterion to take into account simultaneously the spatial coherence of the emission and the multiphase nature of the gas. In order to obtain a solution with spatially smooth parameters, the optimization is performed on the whole data cube at once. The performances of {\tt ROHSA} were tested on a synthetic observation computed from numerical simulations of thermally bi-stable turbulence. An application on a 21\,cm observation of a high Galactic latitude region from the GHIGLS survey is presented.}
{The evaluation of {\tt ROHSA} on synthetic 21\,cm observations shows that it is able to recover the multiphase nature of the HI. For each phase, the power spectra of the column density and centroid velocity are well recovered. More generally that test reveals that a Gaussian decomposition of HI emission is able to recover physically meaningful information about the underlying three-dimensional fields (density, velocity and temperature). The application on a real 21\,cm observation of a high Galactic latitude field produces a picture of the multiphase HI, with isolated, filamentary and narrow ($\sigma \sim 1-2$\,km\,s$^{-1}$) structures and wider ($\sigma \sim 4-10$\,km\,s$^{-1}$), diffuse and space filling components. The test-case field used here contains significant intermediate-velocity clouds that were well mapped out by the algorithm. As {\tt ROHSA} is designed to extract spatially coherent components, it performs well at projecting out the noise.}
{In this paper we are introducing {\tt ROHSA} a new algorithm performing a separation of diffuse sources in hyper-spectral data on the basis of a Gaussian decomposition. The algorithm makes no assumption about the nature of the sources, except that each one as a similar line-width. The tests we made shows that {\tt ROHSA} is well suited to decompose complex 21\,cm line emission of high Galactic latitude regions, but its design is general enough that it could be applied to any hyper-spectral data type for which a Gaussian model is relevant. }

\maketitle
\section{Introduction}
\label{sec::Intro}
Star formation in galaxies is strongly linked to the physical processes that govern the evolution of the interstellar medium.
Stars form by gravitational collapse of dense ($n > 10^4$\,cm$^{-3}$) and cold ($T \sim 10$\,K) structures in molecular clouds
but the process that leads to the formation of these over-densities is still unclear. One key element seems to be related to
the efficiency of the formation of cold clouds of neutral hydrogen (HI) \citep{ostriker_regulation_2010}.

The current vision of the HI comes from an important legacy. Early observations of the 21\,cm line showed a
significant difference between emission and absorption spectra. On lines of sight crossing radio-sources the HI appears
in absorption with very narrow features (a few km\,s$^{-1}$). In emission the 21\,cm line contains these narrow features
on top of much boarder spectral structures (10-20\,km\,s$^{-1}$). \cite{clark_interferometer_1965} was the first to suggest that this
might be the signature of a cloud-intercloud medium in pressure equilibrium. Very rapidly,
\cite{field_thermal_1965,field_cosmic-ray_1969} introduced the concept of thermal instability and laid out the
theoretical ground of a "two phase" HI model showing that, at the pressure of the ISM, the heating and cooling
processes naturally lead to two thermally stable state : a dense cold neutral medium (CNM - $T\sim 50$\,K, $n\sim 50$\,cm$^{-3}$) immersed in
a diffuse warm neutral medium (WNM - $T \sim 8000$\,K, $n \sim 0.3$\,cm$^{-3}$). This vision was later complemented
by \cite{wolfire_neutral_1995, wolfire_neutral_2003} considering updated heating (dominated by the photo-electric effect
on small dust grains) and cooling (dominated by CII - 158 $\mu$m, OI - 63 $\mu$m, L$\alpha$ and electron recombinations
onto positive charged grains) processes of the ISM.

This description of the diffuse neutral gas completed a parallel point of view that emerged in the 50's \citep[e.g.,][]{von_weizsacker_evolution_1951,von_hoerner_methode_1951,chandrasekhar_theory_1952} and that considered the ISM as a multi-scale turbulent medium. In this case the density and velocity structures are the result of a highly dynamical and out-of-equilibrium medium. In order to reconcile the static/two-phase and the turbulent point of views, several studies have aimed at understanding the production of the CNM in a turbulent and thermally unstable flow using numerical simulations \citep[e.g.,][]{hennebelle_dynamical_1999,koyama_origin_2002,audit_thermal_2005,hennebelle_warm_2008,saury_structure_2014}.
In general, these numerical studies show that the WNM has the properties of a trans-sonic turbulent flow, while the CNM
shows a much more contrasted density structure, in accordance with the cloud-intercloud picture.
In addition, such studies indicate the presence of a significant fraction of the mass being in the thermally unstable
regime, (i.e., with a temperature mid-way between the CNM and WNM stable states). For instance, \cite{saury_structure_2014}
showed that 30\% of the HI is in the thermally unstable regime. Interestingly they also show that this lukewarm neutral medium (LNM)
is spatially located around the cold structures, pointing at the transitional nature of this thermal state.

From the observational standpoint, studies combining 21\,cm absorption and emission data have clearly revealed the presence of HI at intermediate/unstable temperatures, typically between 500 and 5000\,K  \citep[e.g.,][]{heiles_millennium_2003,kanekar_temperature_2003,roy_temperature_2013,roy_temperature_2013-1,
murray_21-sponge_2015,murray_21-sponge_2018}. Based on a coherent modeling of emission and absorption spectra \cite{heiles_millennium_2003-1,murray_21-sponge_2015,murray_21-sponge_2018} estimated that about 30\% of the HI is in the cold CNM phase, 20\% in the thermally unstable regime and 50\% in the WNM. Nevertheless the fraction of the HI in each phase remains uncertain and large variations are observed: the fraction of the mass in the CNM ranges from $\sim$1\% to more than 50\% \cite[]{murray_21-sponge_2018}.

The nature of these variations and how they relate to the dynamical conditions of the gas remains largely unexplored from the observational point of view. One main hurdle in getting access to this information is the fact that our knowledge of the multiphase nature of the HI relies on 21\,cm absorption measurements that are limited to lines of sight crossing radio-sources. By nature, this way of observing prevents us from mapping the HI phases. To go further, and really compare with numerical simulation that are, for now, under-constrained by observation, it is mandatory to map the column density structure of each phase and study the spatial variations of their centroid velocity and velocity dispersion. This calls for methods that can extract the information of each HI phase from fully sampled 21\,cm emission data only. Huge efforts have been made to map the 21\,cm emission of the Galactic HI \cite[recent examples are][]
{taylor_canadian_2003,kalberla_leiden/argentine/bonn_2005,stil_vla_2006,mcclure-griffiths_gass_2009,winkel_effelsberg-bonn_2016,peek_galfa-h_2018} and a large amount of data is now available.  The information about the multiphase and multi-scale nature of the HI contained in these large hyper-spectral data cubes has remained elusive due to the difficulty to separate the emission from the different phases on each line of sight. In this paper we propose a new method to map out the contribution of each phase to the 21\,cm emission. The method is based on a decomposition of the 21\,cm emission line with Gaussian profiles and constraints that favors spatially coherent parameters.

This paper is organized as follows. In Sect. \ref{sec::methodology}, we
describe the methodology used to develop our Gaussian decomposition
algorithm. Evaluation on numerical simulation is presented Sect.
\ref{sec::evaluation}. In Sect. \ref{sec::application} we
present an application on observation. The discussion and summary are presented 
Sects. \ref{sec::discussion} and \ref{sec::summary}..
\section{Methodology}
\label{sec::methodology}
\subsection{Gaussian decomposition of the 21\,cm emission}

Very early on after its detection, the 21\,cm line was observed to be well described by a sum of a small number of Gaussian components. This was found to be true for the least confused absorption spectra \citep{muller_21-cm_1957,muller_21-cm_1959,clark_interferometer_1965} but also for emission spectra observed away from the Galactic plane \citep{heeschen_features_1955}. In fact very few spectra at high Galactic latitudes do not comply with that rule whatever the angular resolution of the data. Recently \citet{kalberla_properties_2018} showed that more than 60\% of the spectra over the whole sky can be described by the sum of 7 or less Gaussian components. In the two decades after the detection of the 21\,cm line, many studies used the Gaussian decomposition to infer physical parameters from the data \citep{matthews_observations_1957,davis_21-centimeter_1957,muller_21-cm_1957,muller_21-cm_1959,dieter_neutral_1964,dieter_neutral_1965,lindblad_dwingeloo_1966,takakubo_neutral_1966,mebold_intercloud_1972}.
The fact that a small number of Gaussian components is needed to describe the signal was seen as a convenient way to describe the emission profiles with a small set of parameters \citep{takakubo_neutral_1966}. It is also a very strong element in favor of the Gaussian function as a significant descriptor of the underlying physics. 

\cite{takakubo_neutral_1967} showed that the width of the 21\,cm emission line could be grouped in three components ($\sigma \leq 3$\,km\,s$^{-1}$; $3 < \sigma < 7$\,km\,s$^{-1}$; $\sigma \geq 7$\,km\,s$^{-1}$), a result confirmed later on \citep{mebold_intercloud_1972,haud2007,kalberla_properties_2018}. \citet{takakubo_neutral_1967} also showed that the narrow 21\,cm features are well correlated with Ca+ K line absorption measurements. They concluded that the narrow component, also seen in 21\,cm absorption spectra, are likely to be isolated cold clouds (CNM) in the Solar neighborhood. They also showed that the spatial distribution of the centroid velocity, velocity width and column density of the large feature is compatible with a warm (WNM) and diffuse disk that follows Galactic rotation.
The second group of Gaussians, with a width ($\sigma$) between 3 and 7\,km\,s$^{-1}$, is generally attributed to gas in the thermally unstable range, but a fraction of them could be caused by blending of narrow features.

The exact mass fraction of gas in each phase (CNM, LNM and WNM) is still a matter of debate. In addition, because this knowledge is based on absorption measurements, there is very little information about the structure on the sky of these phases. Being able to separate the different phases on each line of sight would allow us to study the structure and kinematic of the cold phase and its relationship with the more diffuse gas. In theory, one could expect that Gaussian decomposition of the emission spectra could provide such mapping. 

\subsection{Limitation of the Gaussian model}

There are many pitfalls in the description of the 21\,cm emission spectra as a sum of Gaussian components : velocity blending, ambiguities of the number of components, non Gaussian profiles, noise peaks, non-uniqueness of the solution. Another important limitation is the effect of optical depth of the 21\,cm line that modifies the shape of the line. More generally, the main opposition to Gaussian decomposition of emission spectra is that any spectrum can be decomposed that way provided that enough Gaussians are used. If that is the case, how can one be sure that the Gaussian representation provides some real physical information about the emitting gas ? For instance, two spatially disconnected cold structures present on the same line of sight could appear at the same projected velocity. In this case their respective emission profiles would be confused. This effect of velocity blending is affecting both the emission and absorption spectra. It is present in the data, more so at lower Galactic latitudes where the length of the line of sight is larger and the number of HI structures increases. 

This line of reasoning led \citet{dickey_h_1990} to advise against using Gaussian decomposition to analyze 21\,cm spectra. Later on, it continued to be used \citep{verschuur1989,verschuur1994,poppel_warm_1994,haud_gaussian_2000,verschuur2004,begum2010a,martin_ghigls_2015,kalberla_properties_2018} but overall there was a lost of interest, except for the analysis of absorption spectra. Indeed by comparing nearby absorption and emission spectra one can recover the effect of optical depth on the 21\,cm emission. In addition, absorption measurements are only sensitive to cold gas, limiting the velocity blending problem. For these reasons, the Gaussian decomposition continued to be used in this context \citep{dickey_fitting_2003,kanekar_temperature_2003}, and especially after the seminal work of \cite{heiles_millennium_2003-1} who developed a dedicated formalism which was then used in several other studies since \citep[e.g.][]{stanimirovic_thinnest_2005,begum2010a,stanimirovic_cold_2014,lee2015,murray2014,murray_21-sponge_2015,murray_optically_2018}. In fact key information about the nature of the HI come from the joint Gaussian decomposition of emission and absorption spectra. 

\subsection{Development of a new approach}

Following what has been done for the comparison of emission and absorption spectra where the Gaussian decomposition is considered valid, we would like to argue that a similar decomposition could be envisaged for emission data only, at least at high Galactic latitudes where the effect of optical depth of the 21\,cm line has been shown to be negligible \citep{murray_optically_2018}.

The fact that physical information could be obtained using a Gaussian decomposition of absorption data reveals the fact that thermal broadening has a significant effect in shaping the line profile or, in other words, that the dynamics of each HI phase is typical of sub or trans-sonic turbulence. When the amplitude of turbulent and thermal motions are commensurate, the line appears smooth and can be represented by a small number of Gaussian components \citep{miville-deschenes_use_2003}. If the HI at high Galactic latitude is indeed represented by a two phase medium with small, cold and trans-sonic structures immersed in a relatively low Mach number and warm diffuse phase, the Gaussian representation could bear significant physical information. 

 The perspective of mapping the phases of the HI is so important that we ventured to explore new ways of decomposing the emission spectra that could be applicable to the high Galactic sky. The main difficulty is the effect of velocity blending for cold structures. As mentioned by \cite{takakubo_neutral_1966}, there will always be cases where a given spectrum can be fitted with a smaller number of components than its neighbors, if two or more components have similar central velocity and velocity dispersion. One way to avoid this confusion is to look for solutions that have a spatial continuity, or that have a slow spatial variation. \cite{poppel_warm_1994}, \citet{haud_gaussian_2000}, \cite{martin_ghigls_2015} and \cite{miville-deschenes_structure_2017} have implemented Gaussian decomposition methods that uses some information about their neighbors in order to favor spatially coherent solutions. Nevertheless, formally these algorithms do not force solutions to be spatially coherent, they simply provide initial guesses to the fit of a single spectrum based on the most likely solutions found in some area around. The optimization is not bound to this initial guess and it can always converge to another solution that would break the spatial smoothness of the parameter space.

The novelty of the algorithm we present here is that it is the first one that imposes the spatial coherence in the determination of the parameters. In order to do that, all the spectra of the emission cube are fitted at the same time. To make sure that the recovered parameters are spatially smooth, specific regularization terms are added to the cost function with non negativity constraints on the amplitude. This algorithm, called {\tt ROHSA}, is described next.

\subsection{{\tt ROHSA}}
{\tt ROHSA} performs a regression analysis
using a regularized nonlinear least-square criterion. We formulate in this section
the Gaussian model used as well as the energy terms added to the
cost function to take into account simultaneously the spatial
coherence of the emission and the multiphase nature of the gas. The
quasi-Newton algorithm, L-BFGS-B, used to perform the optimization is
then briefly described. Finally, we formulate the algorithm performed by
{\tt ROHSA} based on a multi-resolution process from coarse to fine grid.

\subsubsection{Model}
\label{subsubsec::model}
The data  are the measured brightness temperature $T_B(v_z, \rb)$ at a given projected
velocity $v_z$ across sky coordinates $\rb$. The proposed model
$\tilde T_B\big(v_z, \thetab(\rb)\big)$ is a sum of $N$ Gaussian $G\big(v_z, \thetab_n(\rb)\big)$
\begin{equation}
  \tilde T_B\big(v_z, \thetab(\rb)\big) = \sum_{n=1}^{N} G\big(v_z, \thetab_n(\rb)\big)
  \label{eq::model_gauss}
\end{equation}
with $\thetab(\rb) = \big(\thetab_1(\rb), \dots, \thetab_n(\rb)\big)$ and where
\begin{equation}
  G\big(v_z, \thetab_n(\rb)\big) = \ab_n(\rb) \exp
  \left( - \frac{\big(v_z - \mub_n(\rb)\big)^2}{2 \sigmab_n(\rb)^2} \right)
\end{equation}
is parametrized by $\thetab_n = \big(\ab_n, \mub_n, \sigmab_n\big)$ with
$\ab_{n} \geq \bm{0}$ the amplitude, $\mub_{n}$ the position and
$\sigmab_{n}$ the standard deviation 2D maps of the $n$-th Gaussian
profile across the plan of sky. The residual is
\begin{equation}
  L\big(v_z, \thetab(\rb)\big) = \tilde T_B\big(v_z, \thetab(\rb)\big) - T_B(v_z, \rb)
\end{equation}

The estimated 
parameters $\hat{\thetab}$ are defined as the minimizer of a cost function that includes 
the sum of the squares of the residual
\begin{equation}
  \label{eq:2}
  Q(\thetab) = \frac{1}{2} \, \norm{L\big(v_z, \thetab\big)}_{\bm{\Sigma}}^2 = \frac{1}{2} \, 
  \sum_{v_z, \rb} \left(\frac{L\big(v_z, \thetab(\rb)\big)}{\bm{\Sigma}(\rb)}\right)^2
\end{equation}
where $\Sigma$ is the standard deviation 2D map of 
the noise assumed without spatially correlation. In practice this term is 
estimated using a sequence of empty velocity channels of $T_B(v_z, \rb)$. 

For each of the $N$ Gaussians, we want to obtain a spatially coherent solution, meaning that for each parameter, 
the values have to be close for neighboring lines of sight. This can be done by penalizing the small scale spatial
fluctuations of each parameter, measured by the energy at high spatial frequencies. The considered high-pass filter 
is the second-order differences, that is the Laplacian filtering, defined by the 2D convolution kernel
\begin{equation}
  d = \begin{bmatrix}
  0 & -1 & 0 \\
  -1 & 4 & -1 \\
  0 & -1 & 0 \\
  \end{bmatrix} .
\end{equation}
The following regularization term, containing energy terms is added to 
the cost function given in Eq.~\eqref{eq:2}
\begin{equation}
  \label{eq:3}
  R(\thetab) = \frac{1}{2} \, \sum_{n=1}^N \lambda_{\ab} \|\Db\ab_n\|_2^2 +
  \lambda_{\mub} \|\Db\mub_n\|_2^2 + \lambda_{\sigmab} \|\Db\sigmab_n\|_2^2
\end{equation}
where $\Db$ is a matrix performing the 2D convolution using the kernel $d$ and
$\lambda_{\ab}$, $\lambda_{\mub}$, and $\lambda_{\sigmab}$ are
hyper-parameters than tune the balance between the different terms. 

These terms ensure a positive spatial correlation of the model parameters for neighboring 
pixels. However, each term is free to have large variation across the field at larger 
scale.  
Since $\sigmab_n$ contains information about the gas thermodynamics, we design an additional
term in the cost function to group Gaussians with similar $\sigma_n$. This is implemented 
in order to favour solution that would produce components ascribable to each phase : WNM,
LNM or CNM. In order to do that we add a term, 
$\lambda'_{\sigmab} \|\sigmab_n - m_n\|_2^2$, that constrains $\sigmab_n$ to be close to 
an unknown scalar value $m_n$. The full regularization term is then 

\begin{align}
  \label{eq:4}
  R(\thetab,\mb) = \frac{1}{2} \, \sum_{n=1}^N  & \lambda_{\ab} \|\Db\ab_n\|_2^2 + 
  \lambda_{\mub} \|\Db\mub_n\|_2^2 + \lambda_{\sigmab} \|\Db\sigmab_n\|_2^2 \notag \\ 
  & + \lambda'_{\sigmab} \|\sigmab_n - m_n\|_2^2
\end{align}

\noindent 
with $\mb = (m_1, \dots, m_N)$ and $\ab_n \geq 0$, $\forall \, n \in [1, \dots, N]$.  
The last two terms in Eq.~\eqref{eq:4}, representative of a joined constraint imposed
on $\sigma_n$, allows us to interpret a posteriori and simultaneously the morphology 
and the thermodynamical state of each component extracted from the data. The full cost 
function is then

\begin{align}
  \label{eq:5}
  J(\thetab, \mb) = Q(\thetab) + R(\thetab,\mb)
\end{align}

\subsubsection{Optimization algorithm}
\label{subsubsec::optimization_algorithm}
\label{subsubsec::method}

Unlike $Q(\thetab)$, each energy term proposed in Eq~\eqref{eq:3} involves linear
dependencies on the parameters $\thetab$. The cost function Eq.~\eqref{eq:5} is therefore a regularized non-linear least-square criterion. The minimizer

\begin{equation}
\label{eq:6}
[\hat \thetab, \hat \mb] = \underset{\thetab,\mb}{\text{argmin}}\ J(\thetab,\mb),
\ \text{wrt. } \ab_n \geq 0,\ \forall \, n \in [1, \dots, N]
\end{equation}
as no closed form expression and is not directly tractable because of the complexity of the model
$\tilde{T}_B$ and the size of the unknown and data. The proposed solution relies instead on an iterative optimization algorithm that uses the gradient 

\begin{align}
    \nabla J(\thetab, \mb)
    = \begin{bmatrix}
        \nabla L(\thetab) \times L(\thetab) \\ \bm{0}
    \end{bmatrix} +
    \begin{bmatrix}
        \nabla_{\thetab} R(\thetab,\mb) \\ \nabla_{\mb} R(\thetab,\mb)
    \end{bmatrix} 
\end{align}
which is tractable since it involves the residual, the Jacobian of the residual 
$\nabla L(\thetab)$, and 2D convolutions with the kernel $d$ for $\Db$ and $\Db^t$.
The gradient $\nabla R^t(\thetab, \mb) = [\nabla_{\thetab} R^t(\thetab,\mb), \nabla_{\mb} R^t(\thetab,\mb)]$ and $\nabla J(\thetab)$ are detailed in Appendix \ref{app:gradient}.

For the optimization,  {\tt RHOSA} relies on L-BFGS-B  (for
Limited-memory Broyden–Fletcher–Goldfarb–Shanno with Bounds), a quasi-Newton iterative algorithm  
described by \cite{zhu_algorithm_1997} which allows to take 
into account the positivity constraints of the amplitudes.
In this algorithm, after an initialization $\thetab_{(0)}$, the
solution is approached iteratively by
\begin{equation}
\thetab_{(k+1)} = \thetab_{(k)} - \alpha_{(k)} \bm{H}^{-1}_{(k)} \nabla J\left(\thetab_{(k)},\mb_{(k)}\right)
\label{eq::L-BFGS-B}
\end{equation}
where $\bm{H}^{-1} \nabla J\left(\thetab,\mb\right)$ is approximated with the L-BFGS formula.
The iterations are repeated until one of the two following criteria is met : (1) the total number of evaluations of $J(\thetab,\mb)$ and $\nabla J(\thetab,\mb)$
exceeds a maximum number of iterations defined by the user; (2) the projected gradient is sufficiently small (i.e, |proj $\nabla J(\thetab,\mb)$| / (1+|$J(\thetab,\mb)$|) < 10$^{-10}$).

Due to its non-linearity, the least-square criterion $J(\thetab,\mb)$ described in Eq.~\eqref{eq:5} is likely to 
include local minimizers. Therefore, the L-BFGS-B algorithm used by {\tt ROHSA} is also likely to converge 
toward one of these local minima, making the solution highly dependent on the initialization $\thetab^{(I)}_{(0)}$.  
In order to overcome this difficulty, we designed {\tt ROHSA} on an iterative multi-resolution process from coarse to 
fine grid (described in the following section) to automatically choose $\thetab^{(I)}_{(0)}$ and to converge towards a 
satisfactory local minimum.

\begin{algorithm}[!t]
    \begin{algorithmic}[1]
      \Require{$T_B(v_z, \rb)$, $\thetab^{(0)}$, $\mb^{(0)} = \bm{0}$, $N$, 
      $\lambda_{\ab}$, $\lambda_{\mub}$, $\lambda_{\sigmab}$, 
      $\lambda'_{\sigmab}$}

      \For{$i=1$ to $I$}

      \State \label{sec:tt-rohsa-algorithm1} $\left[\thetab^{(i)},\mb^{(i)}\right] \gets \underset{\thetab,\mb}{\text{argmin}} 
      \ J\left(\thetab^{(i-1)}; \mb^{(i-1)}, \langle T_B \rangle_{i}\right)$. 

      \EndFor

      \Return{$\thetab^{(I)}$, $\mb^{(I)}$}
    \end{algorithmic}
    \caption{{\tt ROHSA} based on a multi-resolution process from
      coarse to fine grid where $\langle T_B \rangle_{i}$ is the
      averaged data at $i$ scale.}
    \label{algo::pseudo-algorithm}
\end{algorithm}

\subsubsection{{\tt ROHSA} algorithm}
\label{subsubsec::algo_rohsa}
\begin{figure}[]
	\centering
    \includegraphics[width=\linewidth]{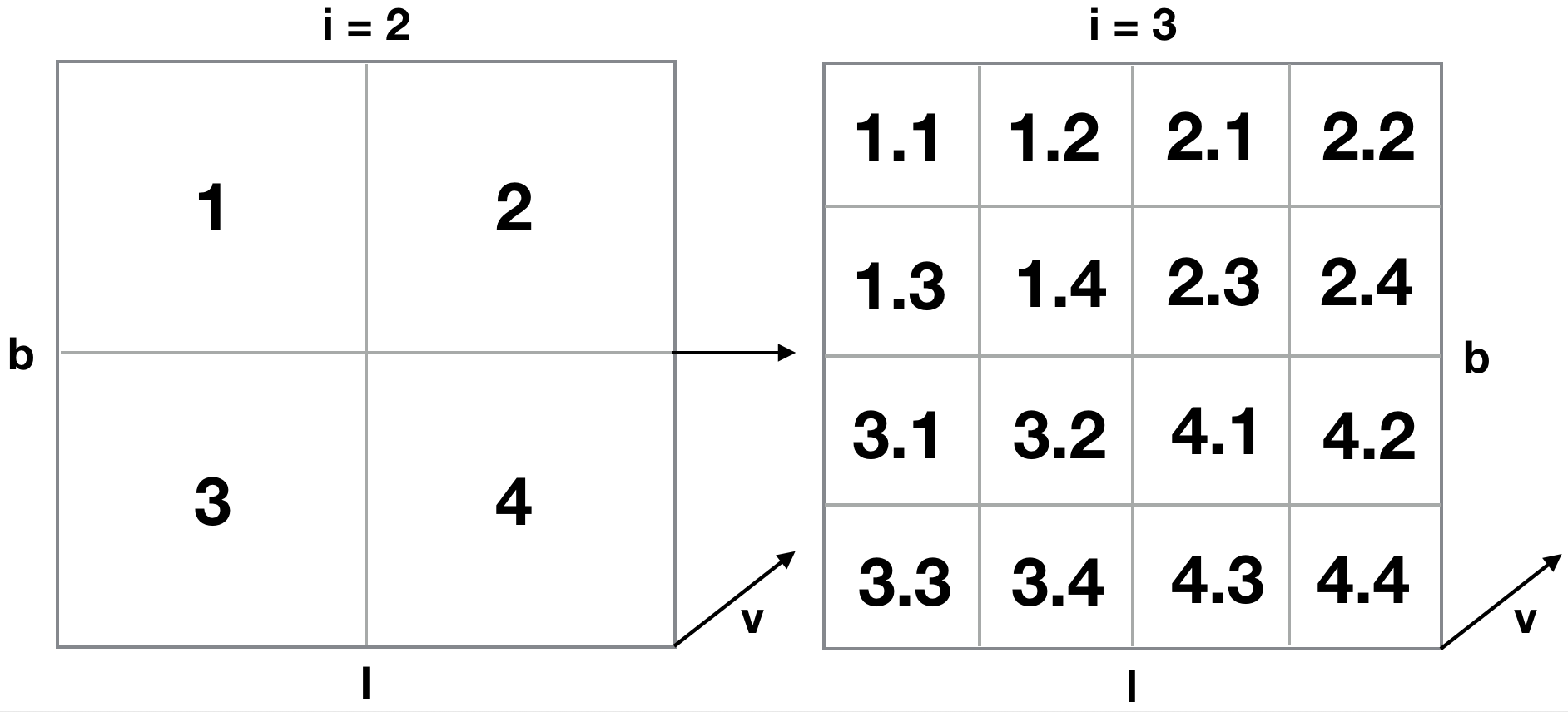}
	\caption{Graphic visualization of neighborhoods $\mathcal{V}_2$
    	and $\mathcal{V}_3$ used to obtain the spatially average
        data versions $\langle T_B \rangle_2$ and $\langle T_B \rangle_3$.}
    \label{fig::main}
\end{figure}

{\tt ROHSA} is based on an iterative algorithm using a
multi-resolution process from coarse to fine grid presented in
Algorithm~\ref{algo::pseudo-algorithm}. The number of iterations $I$ depends on the size of the fine grid $S$ and is defined by the relation $2^I = S$.
For example, a grid of size $S^2$ = $256^2$ requires $I = 8$ iterations.
Each iteration is made of three steps.
\begin{enumerate}
	\item Data are averaged at scale $i$ as
	\begin{equation}
		\langle T_B \rangle_i = \frac{1}{K_i} \sum_{k  \in \mathcal{V}_i} 
    	T_B(v_z, {\rb}_k)
    	\label{eq::mean_Tb}
	\end{equation}
    where $\mathcal{V}_i$ defines the neighborhood at scale $i$, as
    described in Fig.~\ref{fig::main} 
    and $K_i$ is the number of positions in that neighborhood. For $i=1$, all the spatial information is compressed into a single spectrum: $\langle T_B \rangle_1 = \langle T_B(v_z, \rb) \rangle$.
    
    \item The parameters $\thetab^{(i)}$, and $\mb^{(i)}$, are estimated on that
    spatially averaged data version $\langle T_B \rangle_i$ by
    minimizing the cost function given in Eq.~\eqref{eq:5}.

    The minimization (line~\ref{sec:tt-rohsa-algorithm1} of Algorithm~\ref{algo::pseudo-algorithm}) is made using 
    L-BFGS, described in the previous section. Note that
    for scale $i=1$, there is no spatial information
    $\langle T_B \rangle_1$ and the result does not depend on the
    regularization. 
    
	\item Parameters $\thetab^{(i)}$
	are spatially interpolated at nearest neighborhood to serve as initialization 
	for the next scale $i+1$.
\end{enumerate}

The free hyper-parameters $\lambda_{A}$, $\lambda_{\mu}$,
$\lambda_{\sigma}$, $\lambda'_{\sigma}$ remain constant during the iterations. 

\section{Evaluation on numerical simulation}
\label{sec::evaluation}
To evaluate the performance of {\tt ROHSA}, we applied it on synthetic 21\,cm observations computed from a numerical simulation of thermally bi-stable turbulence flow. That allowed us to compare directly the solution given by {\tt ROHSA} to the properties of the gas present in the simulation. That direct comparison with numerical reality is an essential test to evaluate the performances of a source separation algorithm like {\tt ROHSA}. 

\subsection{Numerical simulation}

To test {\tt ROHSA} we used the hydrodynamical simulation of thermally bi-stable turbulence performed by \cite{saury_structure_2014}. We used their 1024N01 simulation ($1024^3$ pixels and a physical size of the box of 40 pc) characterized by
(1) an initial density $n_0 = 0.1$\,cm$^{-3}$, (2) a large scale velocity $v_S = 12.5$\,km\,s$^{-1}$ and (3) a spectral weight $\zeta = 0.2$. The initial density corresponds to the typical density of the WNM before condensation, the large scale velocity represents the amplitude given to the field that generates large scale turbulent motions in the box and finally, the spectral weight controls the modes of the turbulent mixing (here a majority
of compressible modes). The Mach number of this simulation has been evaluated around $\mathcal{M} = 0.85$ for $T > 200$\,K.

In order to explore the performances of {\tt ROHSA} we use only a subset of this simulation. We concentrate our analysis on a $256\times256\times 1024$\,pixels region with moderate CNM fraction in order to limit the effect of HI self-absorption (see Sect.~\ref{sec:optically_thin}).

\begin{figure}[]
  \centering
  \includegraphics[width=0.9\linewidth]{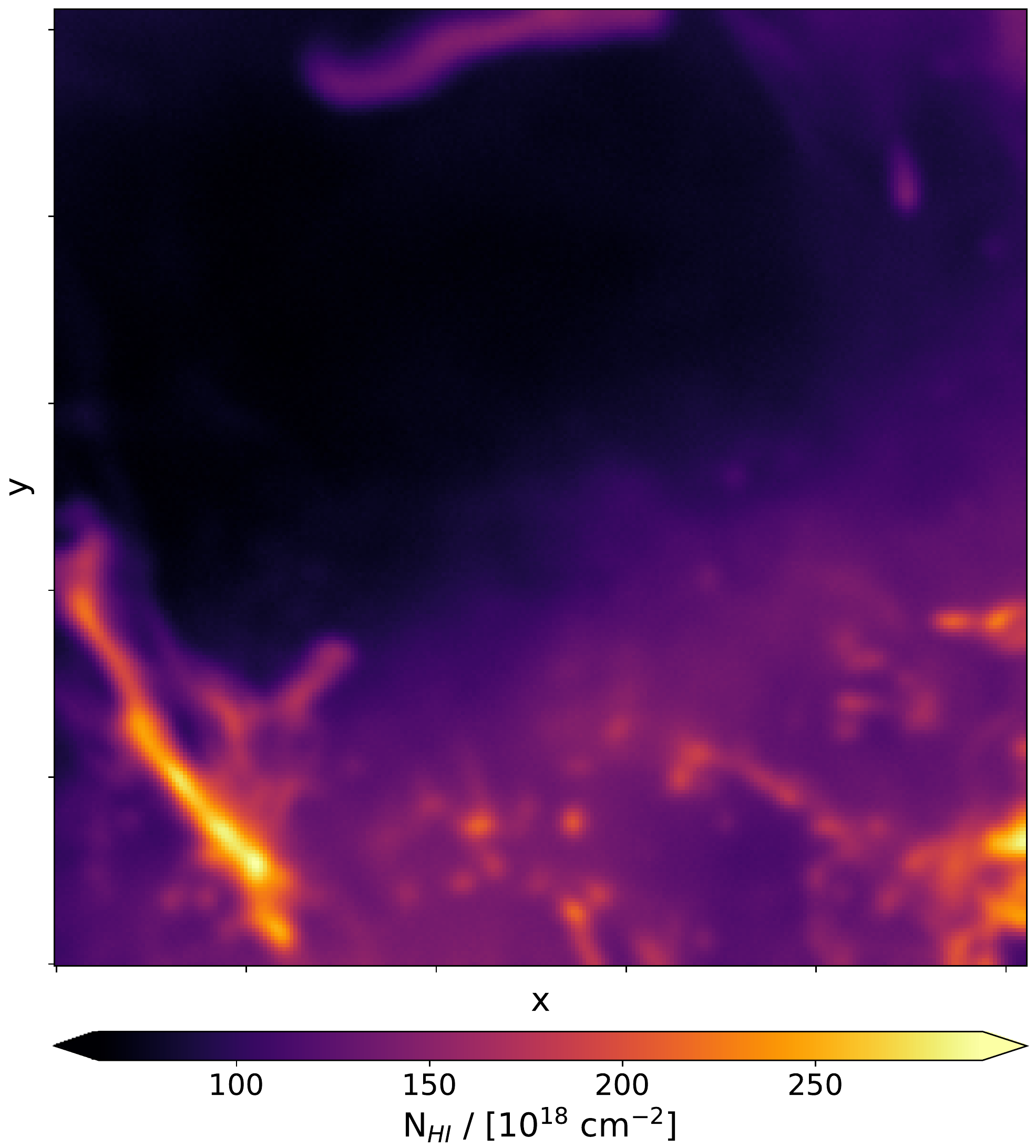}
  \caption{Integrated column density N$_{\rm HI}$ (optically thin approximation) of the 21\,cm synthetic observation computed from the thermally bi-stable numerical simulation of \cite{saury_structure_2014}.}
  \label{fig::NHI_theo_1024}
\end{figure}

\subsection{21\,cm line synthetic observations}
\label{subsec::synthetic_obs}

The synthetic 21\,cm observations were computed 
using the formalism described by \cite{miville-deschenes_physical_2007}. 

\subsubsection{Distribution of velocity fluctuations} 

In the three dimensional spatial space, the neutral hydrogen can be described by three 3D fields : the temperature $T(\textbf{x})$, the density $\rho(\textbf{x})$, and the z-component of the
turbulent velocity field $v_z(\textbf{x})$. Here the three dimensional spatial positions are denoted by the vector \textbf{x} while the two dimensional vector expressing the line-of-sight is denoted by $\rb$. The z-axis is taken along the line of sight. 

Information about the velocity field is inevitably lost because of the projection along z-axis. This make this description of HI a non-exhaustive one. For each position $\textbf{x}$, we assume that the velocity dispersion of a given cell is dominated by thermal motions. This is a fair approximation as the turbulent velocity dispersion at the cell size (0.04\,pc) is $\sigma_{\rm turb} \sim 0.3$\,km\,s$^{-1}$, which is smaller than the thermal broadening everywhere in the simulation: the smallest thermal broadening for the coldest gas found in the simulation ($T=20$\,K) is $\sigma_{\rm therm}=0.4$\,km\,$^{-1}$.
The distribution function of the z-component of the velocity
$v_z(\textbf{x})$ of a given cell is then given by $\phi_{v_z}(\textbf{x})$, a Maxwellian 
shifted by $v_z(\textbf{x})$
\begin{align}
  \phi_{\nu_z}(\textbf{x})dv_z' = \frac{1}{\sqrt{2\pi}\Delta(\textbf{x})} \, \times \, \exp \left( - \, \frac{(v_z' \, - \,
    v_z(\textbf{x}))^2}{2\Delta^2(\textbf{x})} \right)dv_z'
  \label{eq::Maxell_Boltzmann_law}
\end{align} \\
where $\Delta(\textbf{x}) = \sqrt{k_B T(\textbf{x})/m_H}$ is the thermal broadening
of the 21 cm line, $m_H$ the hydrogen atom mass and $k_B$ the Boltzmann constant. 

\subsubsection{Brightness temperature - general case}
\begin{figure}[]
	\centering
    \includegraphics[width=0.9\linewidth]{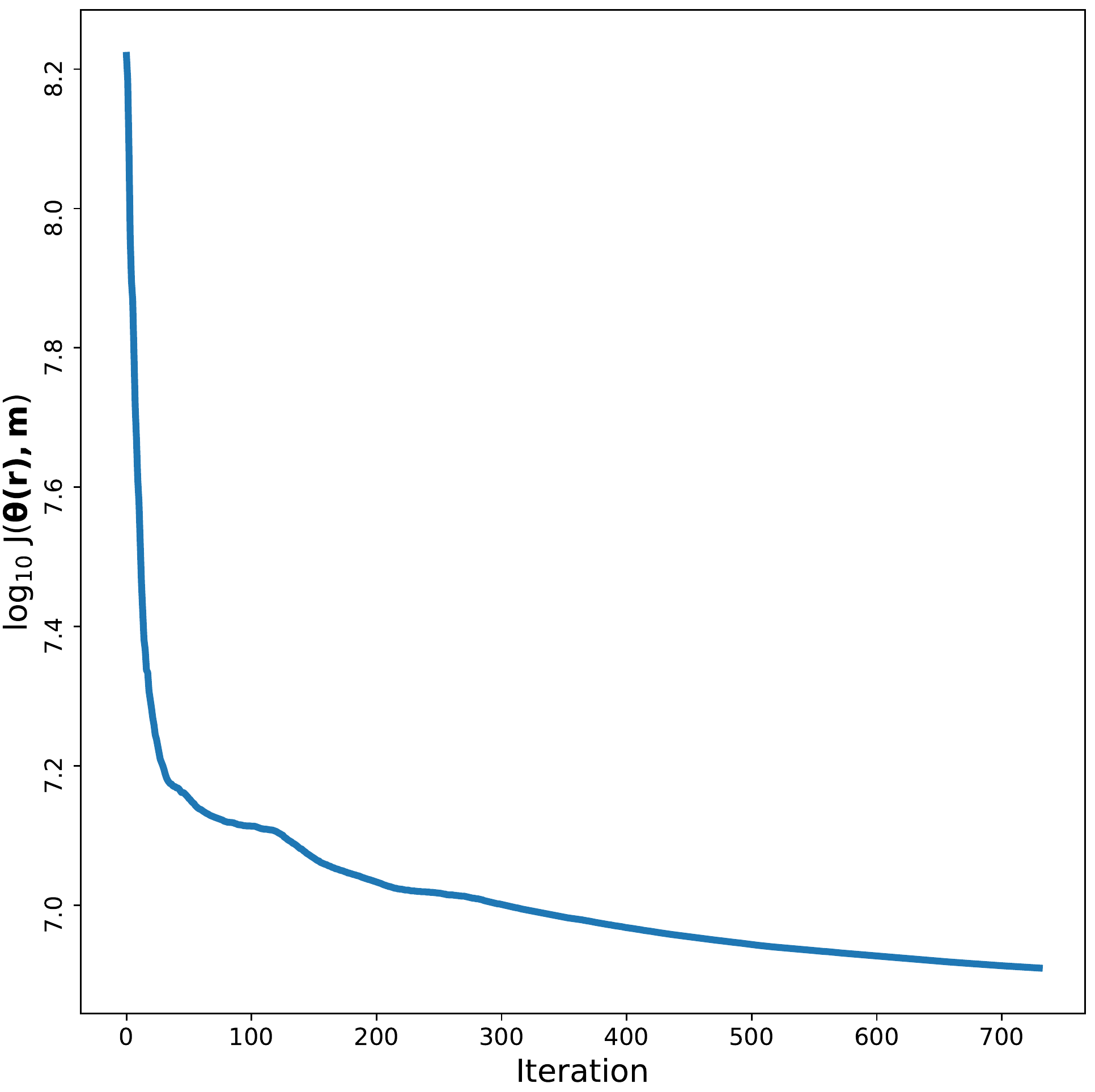}
    \caption{Evolution of the cost function $J(\thetab(\rb),\mb)$ as function of
    the number of iterations performed by {\tt ROHSA} on the synthetic
    observation computed in section \ref{subsec::synthetic_obs}.}
    \label{fig::obf_f}
\end{figure}

The general case for the computation of the 21\,cm  brightness temperature $T_b(v_z, \rb)$ is based on the following radiative transfer equation: 
\begin{align}
	T_b(v_z, \rb) = \sum_z T(\rb,z) \, \left[ 1 - e^{- \tau(v_z,\rb,z)} \right] \,
    e^{- \sum_{z'<z} \tau(v_z,\rb,z')}
	\label{eq::Tb_thick}
\end{align}
where $\tau(v_z,\rb,z)$ is the optical depth of the 21 cm line defined as
\begin{align}
	\tau(v_z,\rb,z) = \frac{1}{C} \, \frac{\rho(\rb,z) \, \phi_{\nu_z}(\rb,z)}
    {T(\rb,z)} \, dz
\end{align}
and C=$1.82243 \times 10^{18}$\,cm$^{-2}$\,(K km s$^{-1}$)$^{-1}$.
In this representation, a gas cell at position $z'$ absorbs emission from cell located behind along the line of sight, i.e., at $z > z'$. 

\subsubsection{Optically thin limit}
\label{sec:optically_thin}

In the optically thin limit, in cases where the self-absorption is negligible (i.e., $\tau(v_z,r,z)<<1$ everywhere), the 21\,cm brightness temperature is proportional to the density $\rho$ :
\begin{align}
  	T_B^{\rm thin}(v_z, \rb) \, dv_z' = B(\rb) \, & \otimes \, \frac{1}{C} \int_0^H dz
    \, \rho(\rb,z) \, \phi_{v_z}(\rb,z) \, dv_z'
    \label{eq::Tb_thin}
\end{align}
where $H$ is the depth of the cloud and $\otimes$ the spatial convolution. Here we consider the case that includes spatial smoothing by a telescope beam $B(\rb)$.

The integrated column density $N_{\rm HI}^{\rm thin}(\rb)$ and the centroid velocity $\langle v_z(\rb) \rangle$ of the 21 cm line can be obtain directly integrating 
$T_B^{\rm thin}(v_z, \rb)$ along the velocity axis :
\begin{align}
  N_{\rm HI}^{\rm thin}(\rb) = C \, \int_{-\infty}^{+\infty} T_B(v_z, \rb)
  \, dv_z
   \label{eq::NHI}
\end{align}
and
\begin{align}
	\langle v_z(\rb) \rangle = \frac{ \int_{-\infty}^{+\infty} \, v_z \, T_B(v_z, \rb)
    \, dv_z}{\int_{-\infty}^{+\infty} \, T_B(v_z, \rb) dv_z}.
    \label{eq::CV}
\end{align}

\subsubsection{Synthetic observation}
\begin{figure}[]
  \centering
  \includegraphics[width=0.9\linewidth]{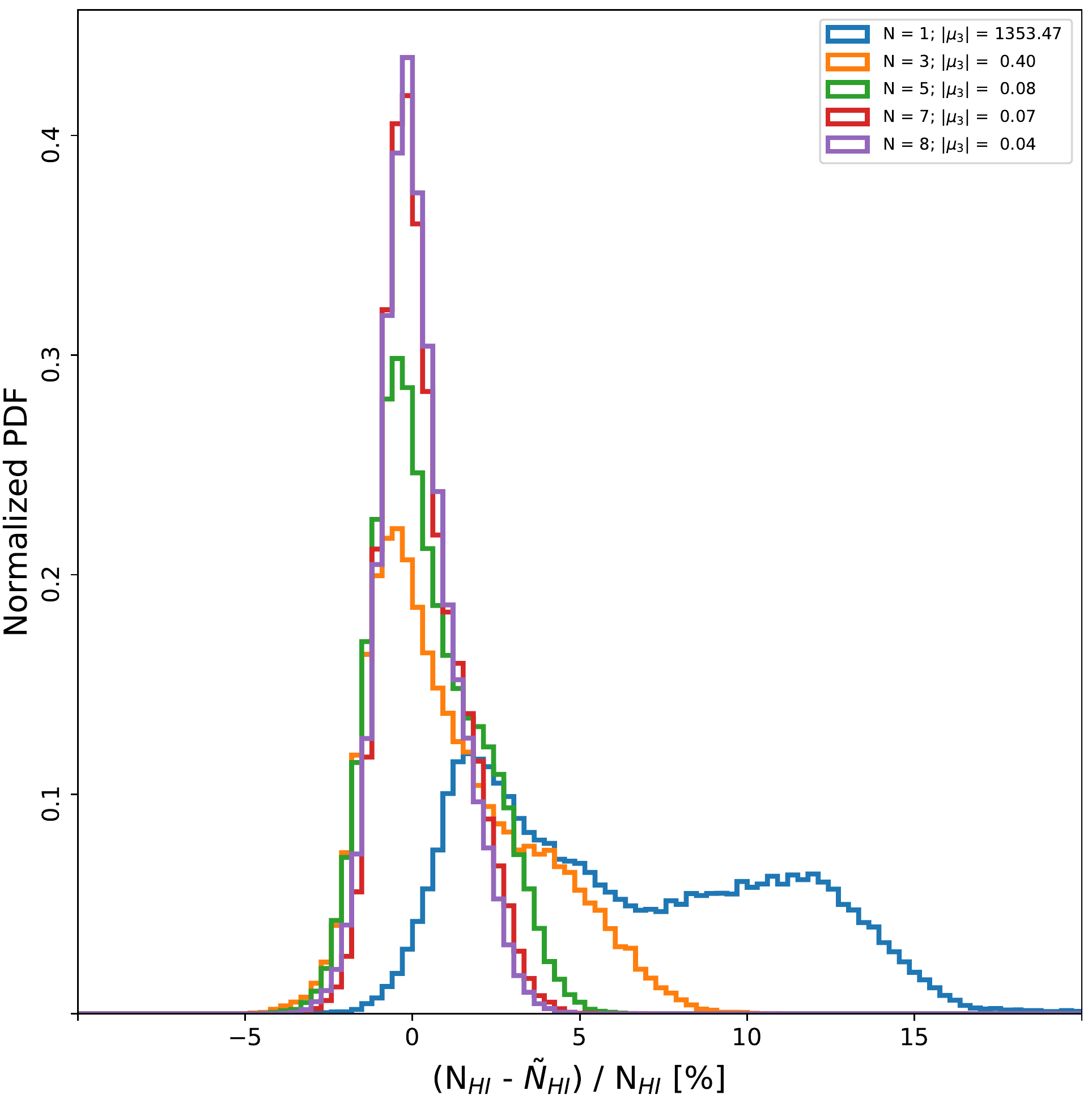}
  \caption{Normalized Probability distribution function of the relative difference (N$_{HI}$ - $\tilde{N}_{HI}$) / N$_{HI}$ between the solution $\tilde{N}_{HI}$ inferred with {\tt ROHSA} and the data N$_{HI}$ for different number of Gaussian components $N = [1,3,5,7,8]$. The norm of the skewness |$\mu_3$| is shown in legend to quantify the quality of the encoding.}
  \label{fig::PDF_residual_SIMU}
\end{figure}

\begin{figure*}[]
  \centering
  \includegraphics[width=\textwidth,height=\textheight,keepaspectratio]{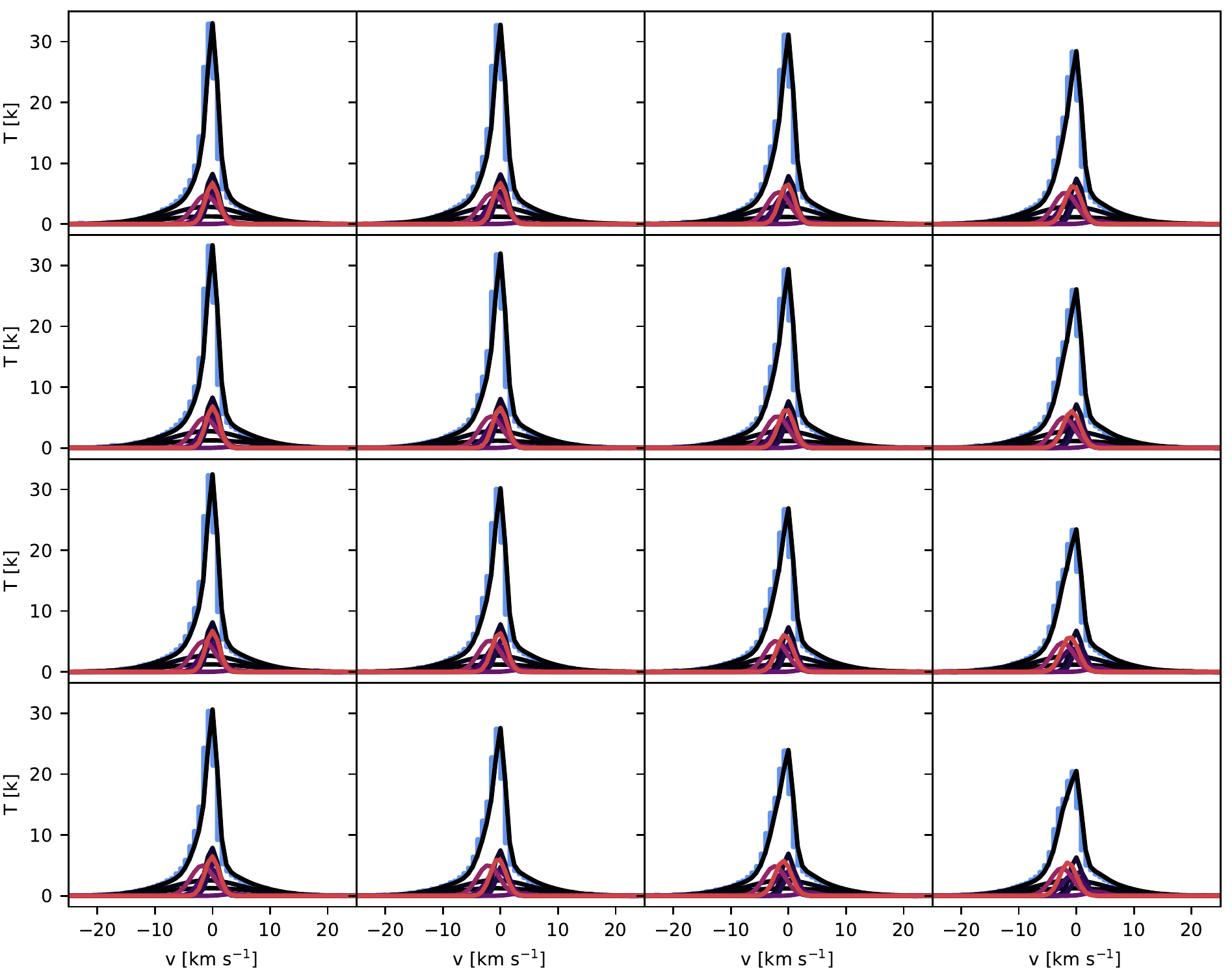}
  \caption{Example of the Gaussian decomposition obtained by {\tt ROHSA}
  for a random 4x4 mosaic of the synthetic observation. The original signal
  is shown in blue and the total brightness temperature encoded by {\tt ROHSA}
  is shown in black. The other lines show the individual Gaussian components. The spatial coherence of the solution can be seen over the mosaic with a smooth variation of the amplitude, the central velocity and the dispersion velocity of each component.}
  \label{fig::mosaic_spectra_all}
\end{figure*}

\begin{table*}[!t]
\centering
\begin{tabular}{l c c c c c c c c}
\toprule
 & $G_{1}$ & $G_{2}$ & $G_{3}$ & $G_{4}$ & $G_{5}$ & $G_{6}$ & $G_{7}$ & $G_{8}$ \\ 
\midrule
$\langle\mub_n\rangle$ [km s$^{-1}$] & 0.2 & -1.7 & 0.5 & 2.3 & -0.1 & 5.0 & -3.7 & -2.5  \\
$\langle\sigmab_n\rangle$ [km s$^{-1}$] & 8.2 & 6.1 & 0.5 & 0.6 & 1.5 & 1.5 & 1.8 & 0.9 \\
\bottomrule
\end{tabular}
\caption{Mean velocity $\langle\mub_n\rangle$ and mean velocity dispersion
	$\langle\sigmab_n\rangle$ of the eight Gaussian components $G_n$ inferred
    by {\tt ROHSA} on the 21\,cm synthetic observation of the numerical simulation described in Sect.~\ref{subsec::synthetic_obs}}
\label{table::mean_var}
\end{table*}

We computed the synthetic Position-Position-Velocity (PPV) data cube in the general case using Eq.~\eqref{eq::Tb_thick}. 
Each spectrum has an effective velocity resolution of 0.8 km s$^{-1}$ and covers 
$-40 < v_z < 40$\,km\,s$^{-1}$.
We considered the beam $B(\rb)$ of the instrument by convolving the synthetic PPV cube with a Gaussian kernel characterized by standard deviations of two pixels along the spatial axis. Then we added a homogeneous Gaussian noise of $0.05$ K to each spectrum. 

In order to mimic observation, integrated column density maps shown in the rest of the paper 
are computed using the optically thin limit presented Eq.~\eqref{eq::NHI}. 
The integrated column density map of the synthetic PPV cube is shown in 
Fig.~\ref{fig::NHI_theo_1024}.

\subsection{Results}
\label{subsec::results}
\begin{figure}[]
  \centering
  \includegraphics[width=0.95\linewidth]{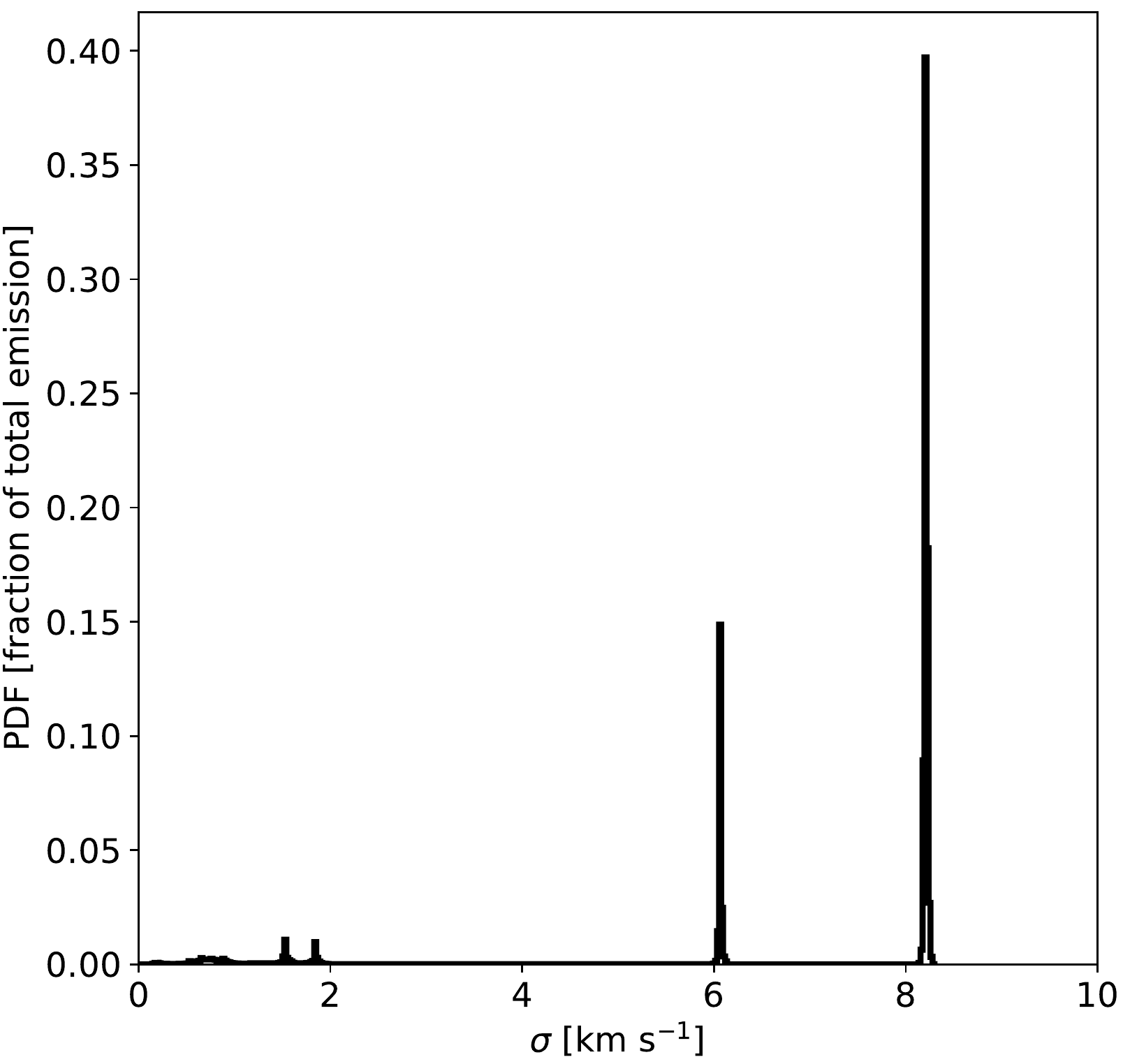}
  \caption{Probability distribution function
  $\sigmab$ weighted by the fraction of total emission of each Gaussian
  $\sqrt{2\pi}\ab_n\sigmab_n/\sum_{\rb}^{}N_{HI}(\rb)$ of the
  simulated field. {\tt ROHSA} converges toward three distinguishable phases
  associated to the WNM, LNM and CNM.}
  \label{fig::PDF_sigma_over_A}
\end{figure}

\begin{figure}[!t]
  \centering
  \includegraphics[width=\linewidth,keepaspectratio]{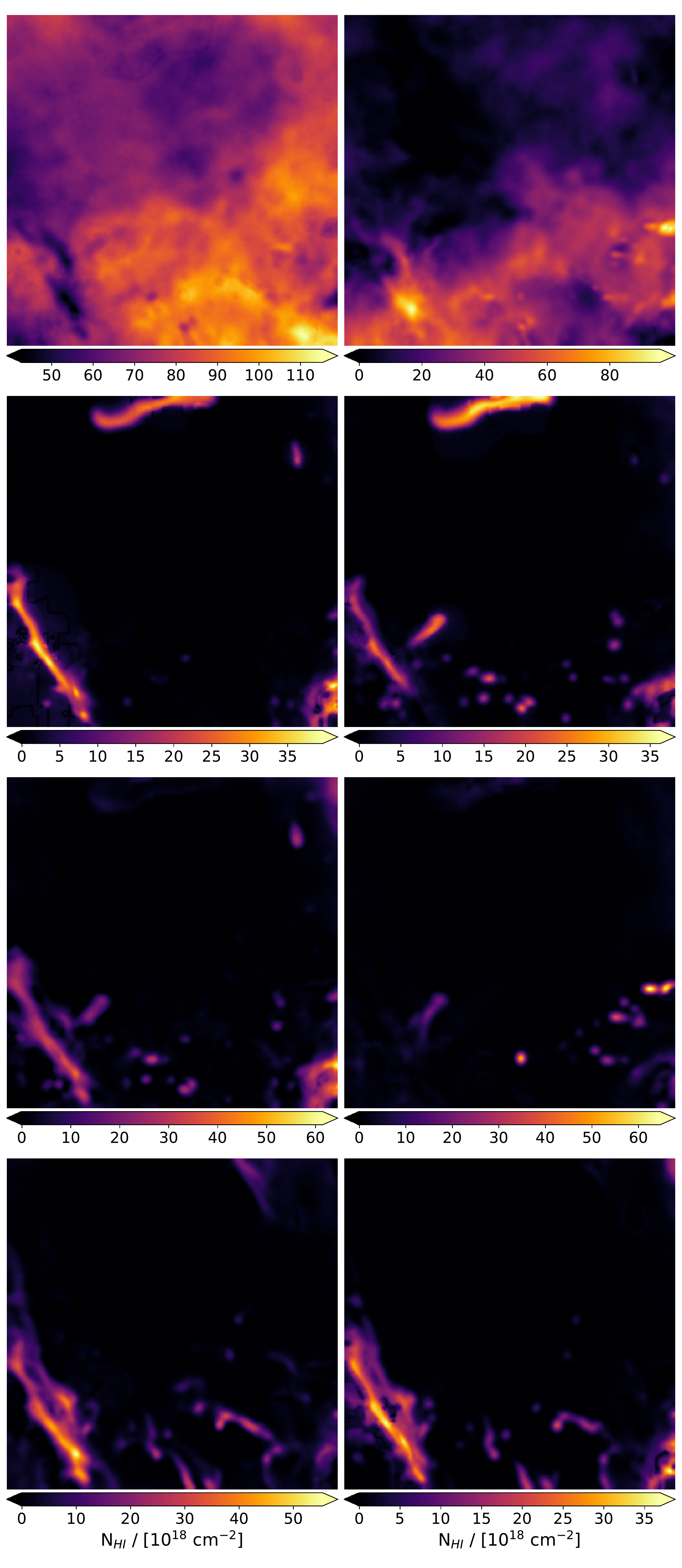}
  \caption{Integrated column density maps (left : $G_1$, $G_3$, $G_5$, $G_7$);
  right : $G_2$, $G_4$, $G_6$, $G_8$) obtained by {\tt ROHSA} on the synthetic
  observation computed in Sect.~\ref{subsec::synthetic_obs}.
  Mean velocity $\langle\mub_n\rangle$ and mean velocity dispersion
  $\langle\sigmab_n\rangle$ are presented in Table~\ref{table::mean_var}.
  The surface filling factor varies considerably between components, depending on their $\langle\sigmab_n\rangle$ value.}
  \label{fig::mosaic_field}
\end{figure}

\begin{figure}[!t]
  \centering
  \includegraphics[width=\linewidth,keepaspectratio]{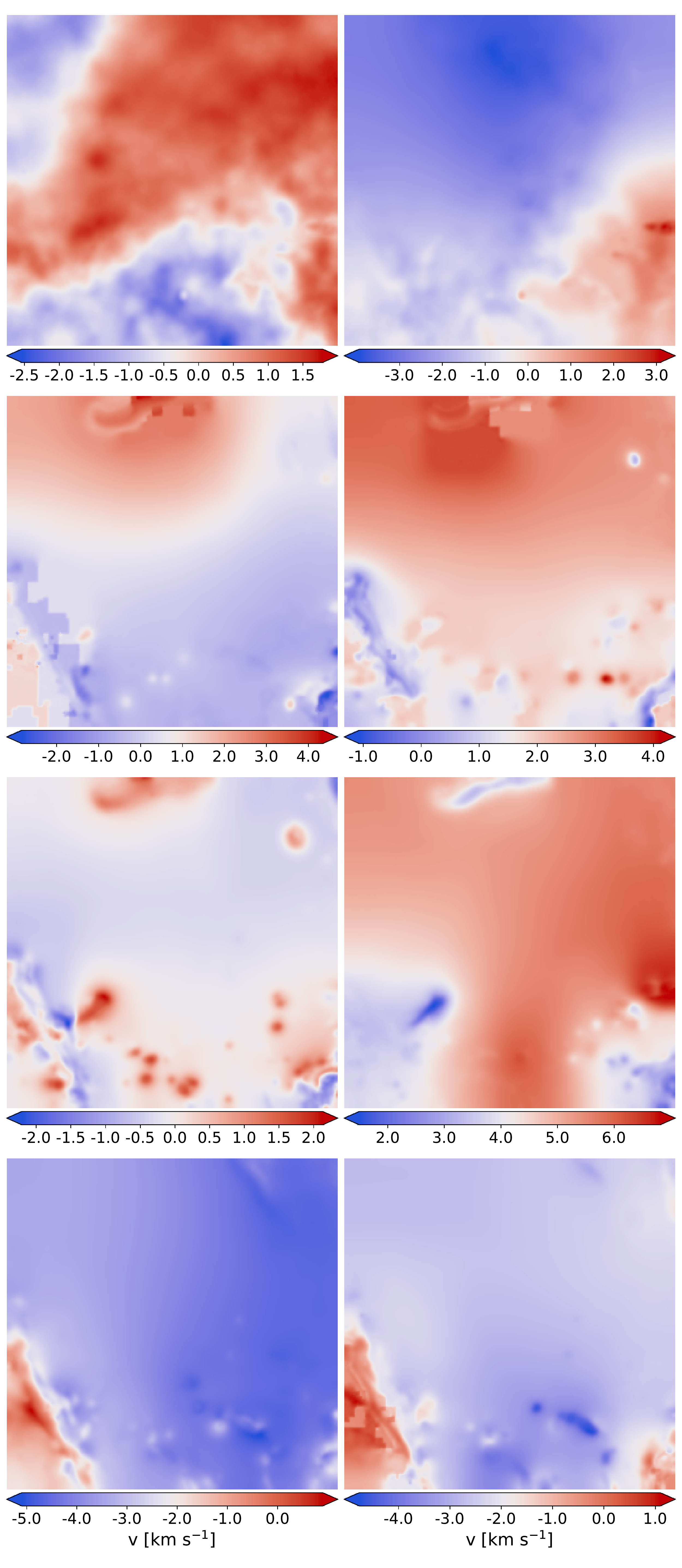}
  \caption{Centroid velocity fields $\mub$ (left : $\mu_1$, $\mu_3$, $\mu_5$, $\mu_7$);
  right : $\mu_2$, $\mu_4$, $\mu_6$, $\mu_8$) obtained by {\tt ROHSA} using the synthetic
  observation computed in Sect. \ref{subsec::synthetic_obs}.} 
  \label{fig::mosaic_vfield}
\end{figure}

\begin{figure}[!t]
  \centering
  \includegraphics[width=\linewidth,keepaspectratio]{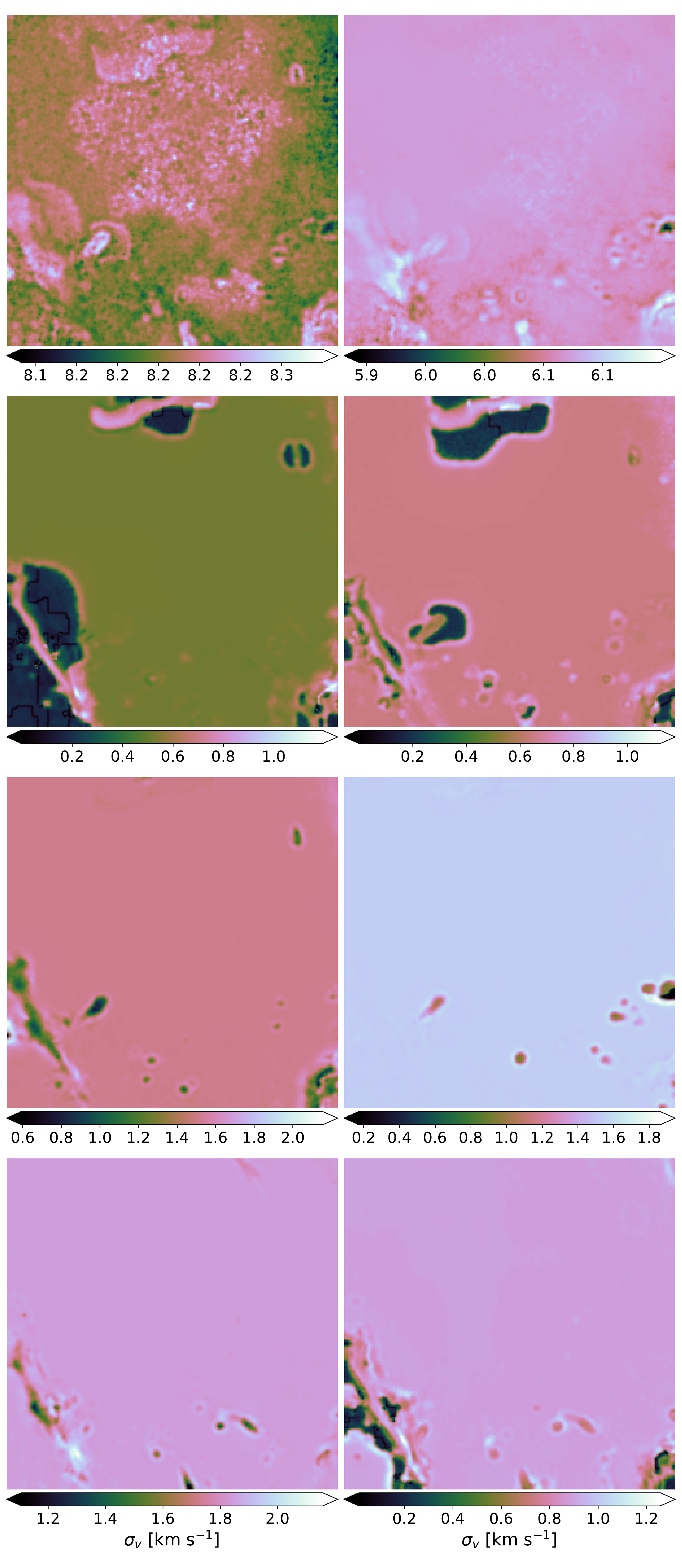}
  \caption{Dispersion velocity fields $\sigmab$ (left : $\sigma_1$, $\sigma_3$, $\sigma_5$, $\sigma_7$);
  right : $\sigma_2$, $\sigma_4$, $\sigma_6$, $\sigma_8$) obtained by {\tt ROHSA} using the synthetic
  observation computed in Sect. \ref{subsec::synthetic_obs}.}
  \label{fig::mosaic_sigfield}
\end{figure}

{\tt ROHSA} was then applied to decompose the synthetic observation computed 
in Sect.~\ref{subsec::synthetic_obs}. 
In this section we discuss the 
choice of the free parameters of {\tt ROHSA}, the global properties of the Gaussian sample and the 
properties of individual components identified by the decomposition.
Then, the mapping of a three-phase coherent model is presented with direct
comparisons to the phases extracted directly from the simulation.

\subsubsection{Choosing the free parameters of {\tt ROHSA}}
\label{subsubsec::free}
{\tt ROHSA} has 6 free parameters : the number of Gaussian components $N$, 4 hyper-parameters $\lambda_i$ and, the maximum number of iteration of the LBFGS algorithm.

The most important parameter is the number of Gaussian $N$. It has to be sufficiently high to ensure a complete encoding of the signal, i.e., that the residual is dominated by noise. As we will discuss in Sect.~\ref{subsubsec::individual}, a given number of Gaussian does not imply that all components are used to describe the signal along every lines of sight. Components are allowed to have an amplitude of zero at any position. This is especially relevant for components encoding cold features. Since the CNM clouds occupy a small fraction of the total volume (see also Sect.~\ref{subsubsec::individual}), we expect the associated amplitude fields to have a large fraction close to zero. This is ensured by the energy term $\|\sigmab_n - m_n\|_2^2$ which minimizes the variance of the dispersion velocity of each component. Amplitudes are brought to zero, if there is no need for a Gaussian to describe the signal at some location, instead of encoding another phase like the WNM for example. Avoiding phase mixing allows to prevent overfitting.

As with the number of Gaussian components $N$, different values of the hyper-parameters
$\lambda_{\ab}$, $\lambda_{\mub}$, $\lambda_{\sigmab}$, $\lambda'_{\sigmab}$ can be tested to obtain a satisfactory solution. If the hyper-parameters are null (i.e., no regularization) the signal could be fully encoded, but no spatial coherence will appear in the solution $\thetab^{(I)}$. On the other hand, if the hyper-parameters are too high, the solution will tend towards a solution that could be too spatially coherent, or even flat, wiping out small scale fluctuations and providing a bad fit to the data. A spatially coherent solution that describes the data well with the smallest value of $N$ is a good criterion to select the values of the hyper-parameters.

The last criterion set by the user is the maximum number of iterations of the LBFGS algorithm computing Eq.~\eqref{eq::L-BFGS-B} (see section \ref{subsubsec::optimization_algorithm}). That parameter must be large enough to ensure the convergence of the solution. The convergence of {\tt ROHSA} from a numerical perspective is presented in Fig.~\ref{fig::obf_f} that shows the evolution of the cost function $J(\thetab(\rb),\mb)$ for 730 iterations.

The decomposition of the synthetic observations presented in this section converged to a satisfactory solution with $N = 8$, $\lambda_{\ab}$ = 10000, $\lambda_{\mub}$ = 10000, $\lambda_{\sigmab}$ = 10000, and $\lambda'_{\sigmab}$ = 1000. As previously recommended, these values are empirically found to converge towards a noise-dominated residual and a signal that is encoded with a minimum number of Gaussian. To illustrate this, Fig.~\ref{fig::PDF_residual_SIMU} shows the normalized probability distribution function of the relative difference (N${HI}$ - $\tilde{N}{HI}$) / N${HI}$ between the solution $\tilde{N}{HI}$ inferred with {\tt ROHSA} and the data N$_{HI}$ for different number of Gaussian components $N$.

As a posterior assessment, the skewness $\mu_3$ of the residual is shown to quantify the quality of the encoding. As the emission is only positive and noise has a symmetric distribution centered on zero, the skewness of the residual can be used as a way to evaluate if the emission is well estimated or if the algorithm has over-fitted the data and included some noise in the solution. A positive skewness is usually an indication that some emission is left in the residual. The model has not enough freedom to encode the emission fully; more components are needed or the regularization terms should be lowered. On the other hand a negative skewness indicates that the decomposition is over-fitting the data; positive noise fluctuations are included in the model leaving more negative noise fluctuations than positive ones in the residual. This is usually an indication that the regularization coefficients ($\lambda_i$) should be larger to increase the spatial smoothness of the solution. A skewness of the residual close to zero is an indication of a valid solution, one that is not distorted by the regularization and that does not overfit the signal. For the case of the synthetic observations presented here, $N = 8$ fully encodes the signal with a relatively low skewness |$\mu_3$| = 0.04.

The computation time of {\tt ROHSA} depends on the maximum number of iterations, the dimension of the PPV cube and the number of Gaussians. For each step of the multi-resolution process from coarse to fine grid, the computation time used by {\tt ROHSA} for this particular case (purple line) is presented in Appendix \ref{app::timestep} with lower number of Gaussian for the sake of comparison. The evaluation performed here requires nearly two hours of computation time on a single CPU, which makes it difficult to explore a large range of hyper-parameters and number of Gaussian. To overcome this difficulty, A GPU implementation of {\tt ROHSA} is under development.

\begin{figure}[!t]
  \centering
  \includegraphics[width=\linewidth,keepaspectratio]{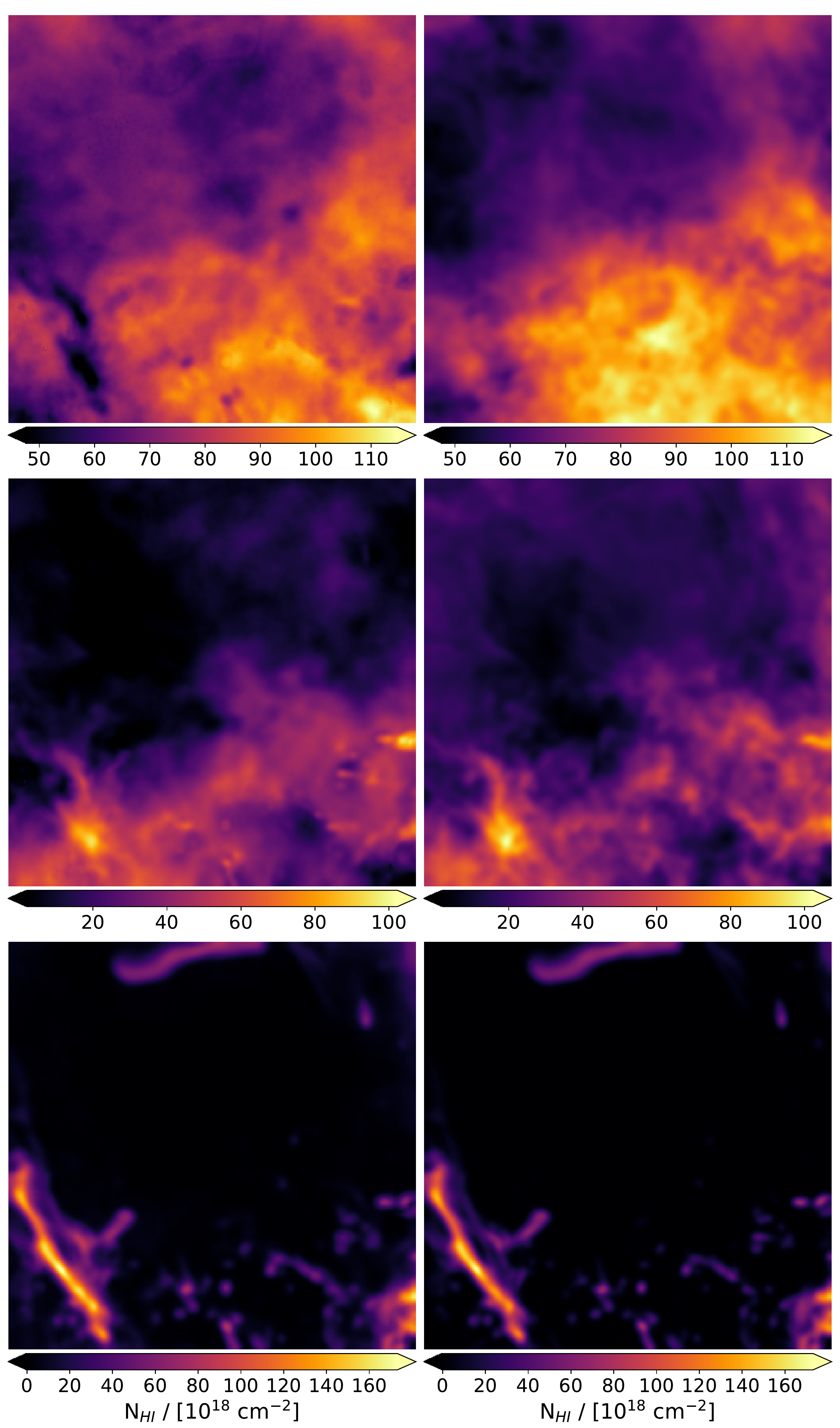}
  \caption{Left : Integrated column density maps of the three-phase model extracted by {\tt ROHSA}. Right : Integrated column density maps of the three-phase model inferred directly from the simulation using the canonical values $T_{k \, {\rm lim,CNM/LNM}}$ = 500 and $T_{k \, {\rm lim,LNM/WNM}}$ = 5000 K. WNM, LNM and CNM are presented from the top to the bottom.}
  \label{fig::mosaic_field_comparison}
\end{figure}

\subsubsection{Global properties of the Gaussian sample}
\label{subsubsec::global}
{\tt ROHSA} recovered the total emission of the synthetic observation with
a relative variation of 0.3 \%.
An example of the Gaussian decomposition for a representative 4x4 mosaic of the
simulation is shown Fig.~\ref{fig::mosaic_spectra_all}. The spatial
coherence of the solution can be seen over the mosaic with a smooth
variation of the amplitude, the central velocity and the velocity dispersion of each
Gaussian. It is already possible to distinguish in those spectra the convergence
of the decomposition toward different velocity dispersions, i.e different
temperatures/phases of the gas due to the energy term $\lambda'_{\sigmab} \|\sigmab_n - m_n\|_2^2$.
To have a clear view of these different components,
let's take a look at the probability distribution function
$\sigmab$ weighted by the fraction of total emission of each Gaussian $\sqrt{2\pi}\ab_n\sigmab_n/\sum_{\rb}^{}N_{HI}(\rb)$ presented in Fig.~\ref{fig::PDF_sigma_over_A}.
This diagram shows the amount of gas in a given
range of velocity dispersion (i.e indirectly a certain range of temperature).
It is clear that {\tt ROHSA}, in this case, converges toward a three-phase model with typical velocity dispersion close to the expected values in the CNM ($\sigma < 2$\,km\,s$^{-1}$), LNM ($\sigma \sim 6$\,km\,s$^{-1}$) and WNM ($\sigma \sim 8$\,km\,s$^{-1}$). 
We will see that a similar behavior is also present in the application to a high Galactic latitude observation presented in Sect. \ref{sec::application}. Note also that since eight Gaussian have been used 
by {\tt ROHSA}, some phases are encoded by several components. Association of the different 
components to characterize a three-phase model will be presented in Sect.~\ref{subsubsec::mapping}.

\subsubsection{Properties of individual components}
\label{subsubsec::individual}
\begin{figure}[]
  \centering
  \includegraphics[width=\linewidth]{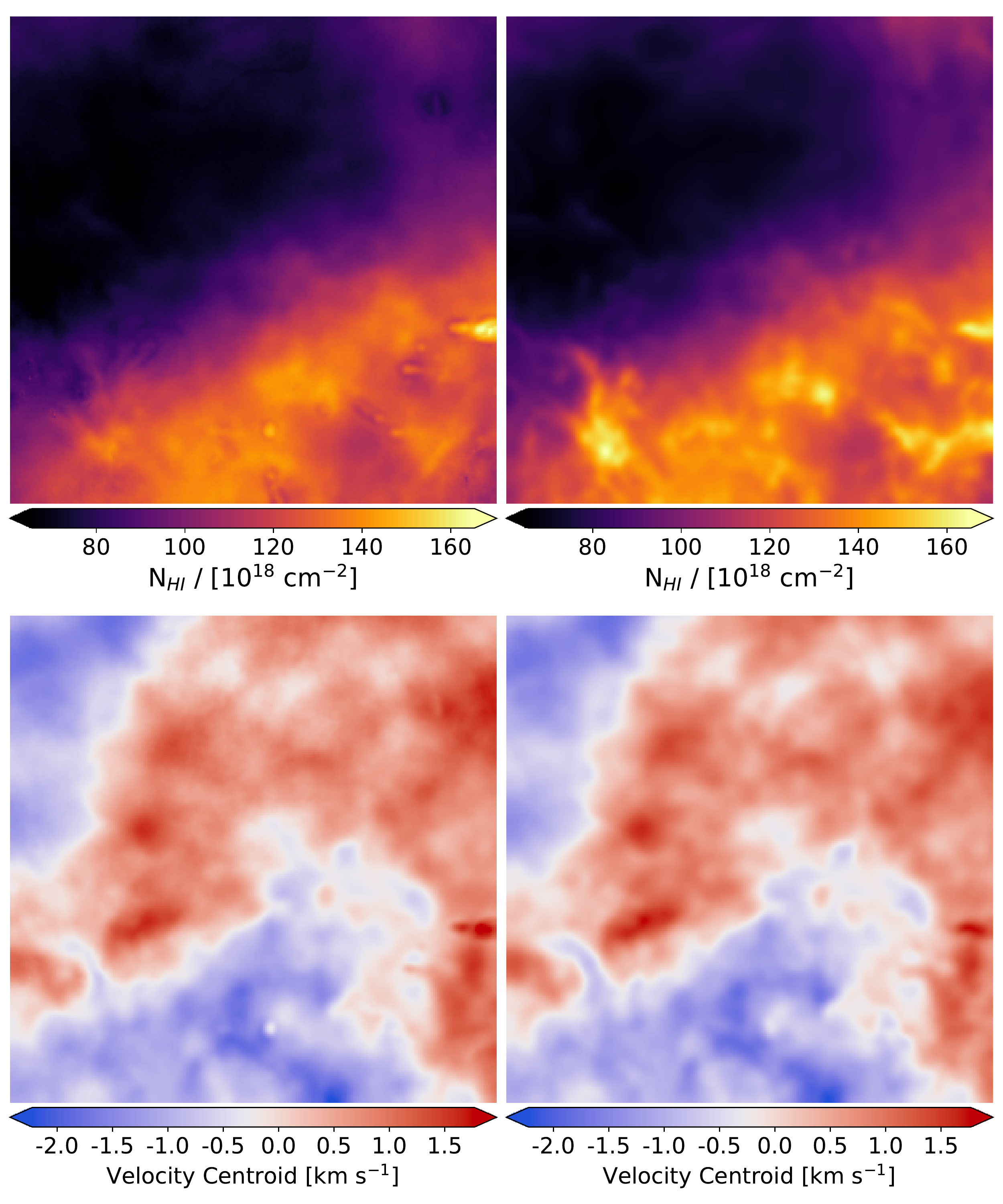}
  \caption{Synthetic observation of the integrated column density field (top) and the 
  centroid velocity field (bottom) associated to the inter-cloud medium (WNM+LNM). Left : Inferred with {\tt ROHSA}; Right : inferred directly from the simulation using all cells with 
  $T_{k \, lim,LNM/WNM}$ > 500 K. 
  For statistical comparison, the spatial power spectra of each one are shown 
  in Figs. \ref{fig::sps1d_NHI} and \ref{fig::sps1d_CV_INTER}.}
  \label{fig::NHI_CV_INTER}
\end{figure}

Integrated column density maps of each components obtained with {\tt ROHSA} for $N = 8$ Gaussian are presented in Fig.~\ref{fig::mosaic_field}. For each component $G_n$, the mean velocity $\langle\mub_n\rangle$ and mean velocity dispersion $\langle\sigmab_n\rangle$ averaged over the field are presented in Table \ref{table::mean_var}. The surface filling factor appears to vary considerably between the eight components. The components with low values of $\langle\sigmab_n\rangle$ are sparsely present, while the component with the largest velocity dispersion is present everywhere on the field. 

We recall that the numerical simulations used here were designed to reproduce the WNM-CNM condensation process of the HI through the thermal instability. Because of the factor 100 difference in density between the two phases, and the fact that the mass fraction in each one is about 50\%, it implies that the cold phase fills only a few percents of the volume \citep{saury_structure_2014}. This translates directly in the column density maps recovered by {\tt ROHSA}; the narrow components, corresponding to colder structures, fill only a fraction of the projected field of view while the larger component is present everywhere. 

The eight velocity fields and velocity dispersion fields are presented in Figures~\ref{fig::mosaic_vfield} and \ref{fig::mosaic_sigfield} respectively. At some location of the fields, when there is no need for a Gaussian to describe the signal over several pixels, the amplitude goes to zero. The corresponding velocity and velocity dispersion fields have then no reason to fluctuate. It turns out that where $a_n$ goes to zero, $\mub_n$ and $\sigmab_n$ are flat. 
It explains the apparent variation of spatial resolution as seen for example  in Fig.~\ref{fig::mosaic_vfield} (bottom right). We notice that the first component $G_1$ (top left) and the second component $G_2$ (top right), encoding respectively the WNM and the LNM, are defined everywhere, so $\mub_1$ and $\sigmab_1$ also have fluctuations everywhere. Implications will be discuss in Sect.~\ref{sec::discussion}.   

\begin{figure}[]
  \centering
  \includegraphics[width=0.95\linewidth]{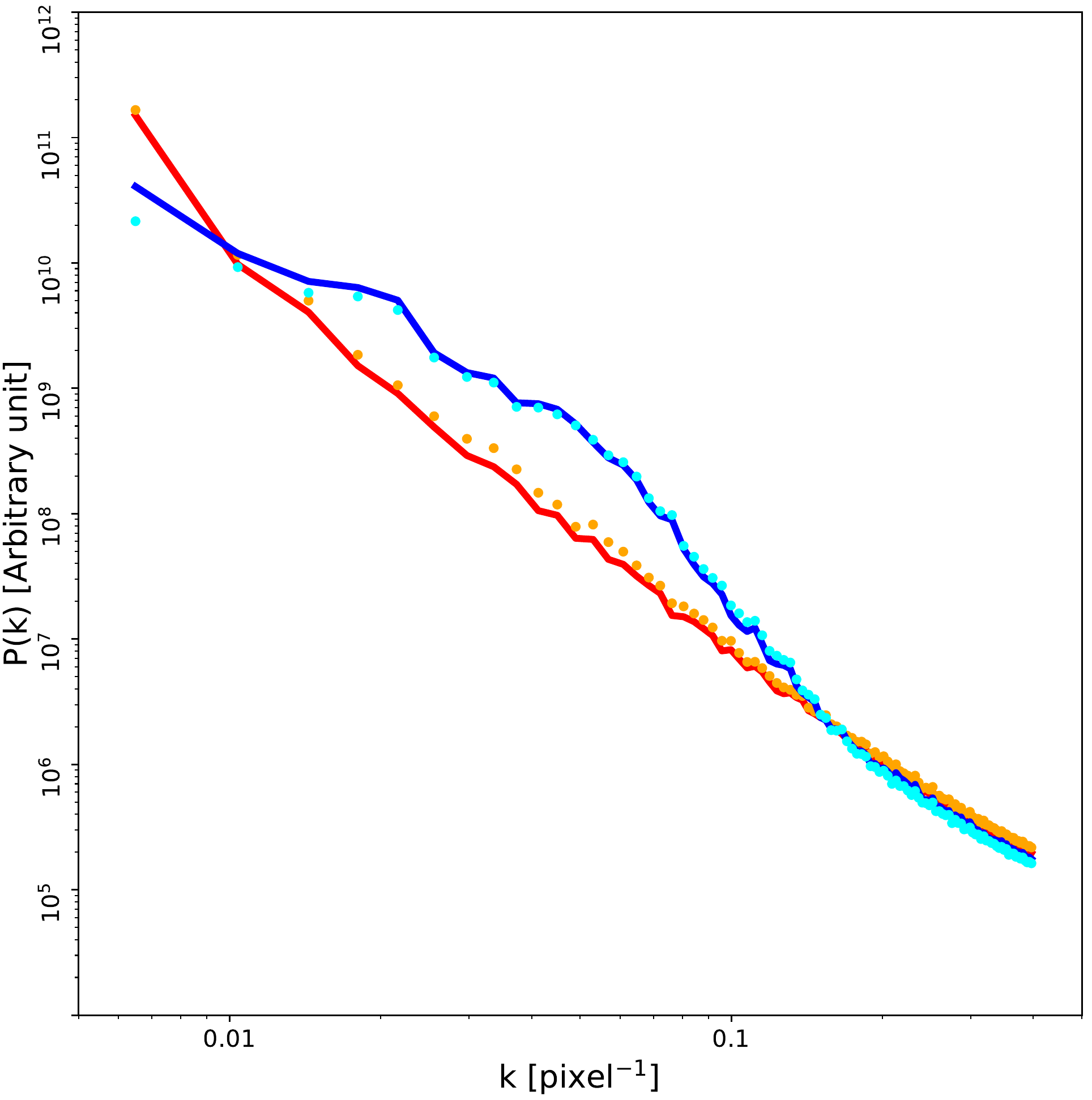}
  \caption{Spatial power spectrum of the column density.  The inter-cloud medium (WNM+LNM) is represented by the orange dotted line (simulation) and the red line ({\tt ROHSA}). The CNM is shown as cyan dotted line (simulation) and blue line ({\tt ROHSA}).}
  \label{fig::sps1d_NHI}
\end{figure}

\subsubsection{Mapping the three-phase neutral ISM}
\label{subsubsec::mapping}
In order to compare the result of the decomposition with the reality given by the numerical simulations, we grouped the eight components into three fields corresponding to the WNM, the LNM and the CNM. The comparison with the numerical simulations requires that we identify ranges in temperature that demarcate these three phases. As with most numerical simulations that include the classical heating and cooling processes of the ISM \citep{wolfire_neutral_1995}, the simulation of \cite{saury_structure_2014} we use here shows a continuum of temperature, with no clear separation and with a significant fraction of the gas present at temperatures corresponding to the thermally unstable regime (see their Figures 14 and 15). To facilitate the comparison with previous studies, we decided to use the canonical values $T_{k \, {\rm lim,CNM/LNM}}= 500$\,K and $T_{k \, {\rm lim,LNM/WNM}}= 5000$\,K \citep{heiles_millennium_2003-1} to separate the simulation in three components. The integrated column density maps associated to each phase are computed following the methodology described in Sect.~\ref{subsec::synthetic_obs}.

The comparison between the integrated column density maps recovered with {\tt ROHSA} and those inferred directly from the simulation is presented Fig.~\ref{fig::mosaic_field_comparison}. The intensity and the morphology of each phase is well recovered. It is nevertheless possible to see some leakages between the phases, in particular between WNM and LNM. This is due partly to the not well defined temperature thresholds used to separate the phases. It is also due to small confusions during the Gaussian decomposition where the intensity, the dispersion velocity and velocity centroid of each component in the PPV space are close to each other. In other words, for similar central velocities, the scales of fluctuations on the velocity axis characterizing each component are too close (see Figs. \ref{fig::mosaic_spectra_all} and \ref{fig::PDF_sigma_over_A}). 

\begin{figure}[]
  \centering
  \includegraphics[width=0.95\linewidth]{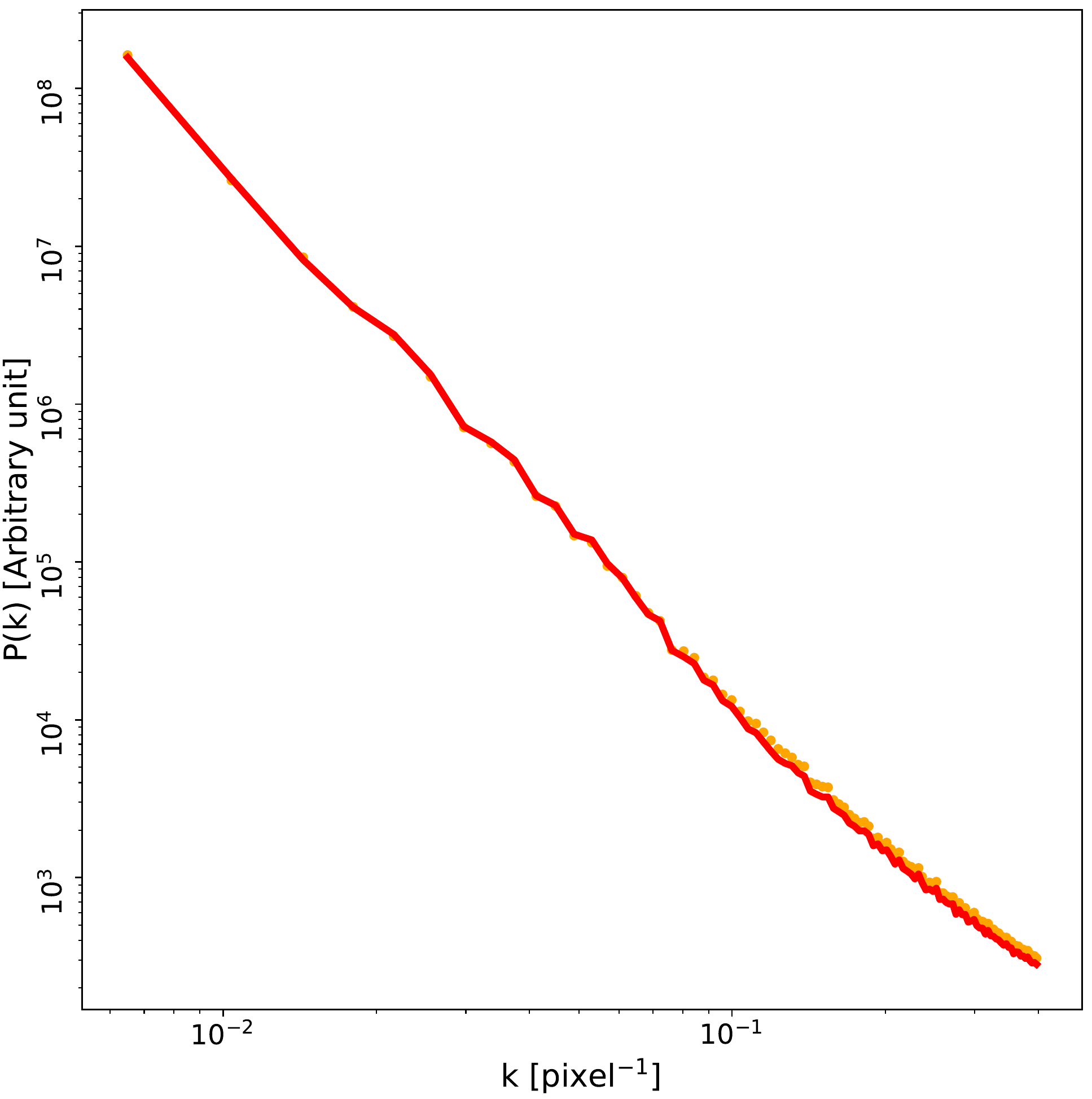}
  \caption{Spatial power spectrum of the centroid velocity field for the inter-cloud medium (WNM+LNM). Orange dot line : inter-cloud medium inferred directly from the simulation using all cells with $T_{k \, lim,LNM/WNM}$ > 500 K (bottom right Fig. \ref{fig::NHI_CV_INTER}). Red line : inter-cloud medium inferred with {\tt ROHSA} (bottom left Fig. 
  \ref{fig::NHI_CV_INTER}).
  }  
  \label{fig::sps1d_CV_INTER}
\end{figure}

\begin{figure*}[!t]
  \centering
  \includegraphics[width=\textwidth,height=\textheight,keepaspectratio]{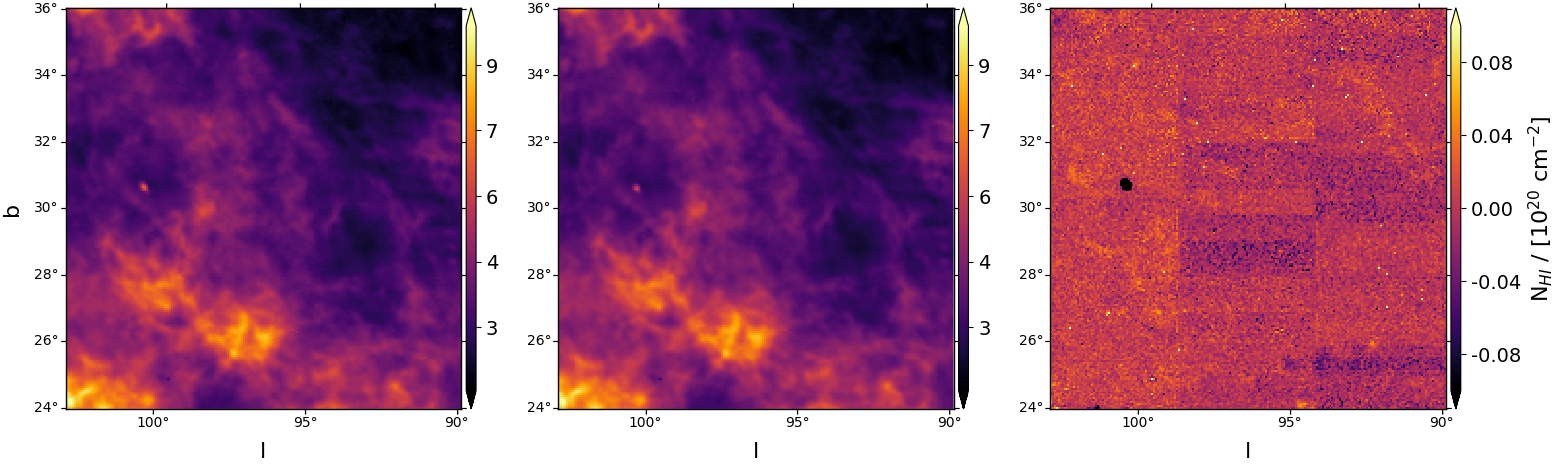}
  \caption{From left to right: Integrated column density $N_{\rm HI}$ of the NEP field which is part of the GHIGLS survey. $N_{\rm HI}$ was computed in the optically thin approximation (see Eq.~\ref{eq::NHI}); Integrated column density $\tilde{N}_{\rm HI}$ of NEP inferred with {\tt ROHSA}; Residual $\tilde{N}_{\rm HI}$-$N_{\rm HI}$ between the model and the data.}
  \label{fig::NHI_TOT_NEP}
\end{figure*}

One way to evaluate the quality of the reconstruction is to compare the statistical properties of the cloud and inter-cloud components. 
Three different fields are used : the integrated column density field of the cloud medium (CNM), the integrated column density field of the inter-cloud medium (LNM+WNM), and because it is fully sampled in the plan of sky, the centroid velocity field of the inter-cloud medium. 
Figure \ref{fig::NHI_CV_INTER} presents the integrated column density field and the centroid velocity field, computed using Eq.~\eqref{eq::CV}, of the inter-cloud medium obtained combining the LNM and the WNM inferred with {\tt ROHSA} and those obtained directly from the simulation. 

In order to compare the estimates from {\tt ROHSA} to the ones obtained from the simulation we compute the spatial power spectrum (SPS) of each image. 
The SPS of the integrated column density of the cloud and inter-cloud medium are presented Fig.~\ref{fig::sps1d_NHI} and the SPS of the centroid velocity field of the inter-cloud medium is presented Fig.~ \ref{fig::sps1d_CV_INTER}. In each case, the statistics recovered by ROHSA is consistent with the numerical simulation over all scales. The shape of these power spectra is interesting in itself; the inter-cloud medium is featureless with an almost constant power law, as the cloud phase is more structured, with a break at about 20 pixels, showing a typical scale linked to the condensation process. Interestingly, {\tt ROHSA} is able to capture all these features very well. 

\section{Application on high-latitude HI gas}
\label{sec::application}

After validating the identification of the HI phases on numerical simulations, in this section we present the application of {\tt ROHSA} on a 21\,cm observation of a high Galactic latitude region. 

\subsection{North ecliptic pole (NEP)}
\label{subsec::NEP}
\begin{figure*}[!ht]
  \centering
  \includegraphics[width=\textwidth,height=\textheight,keepaspectratio]{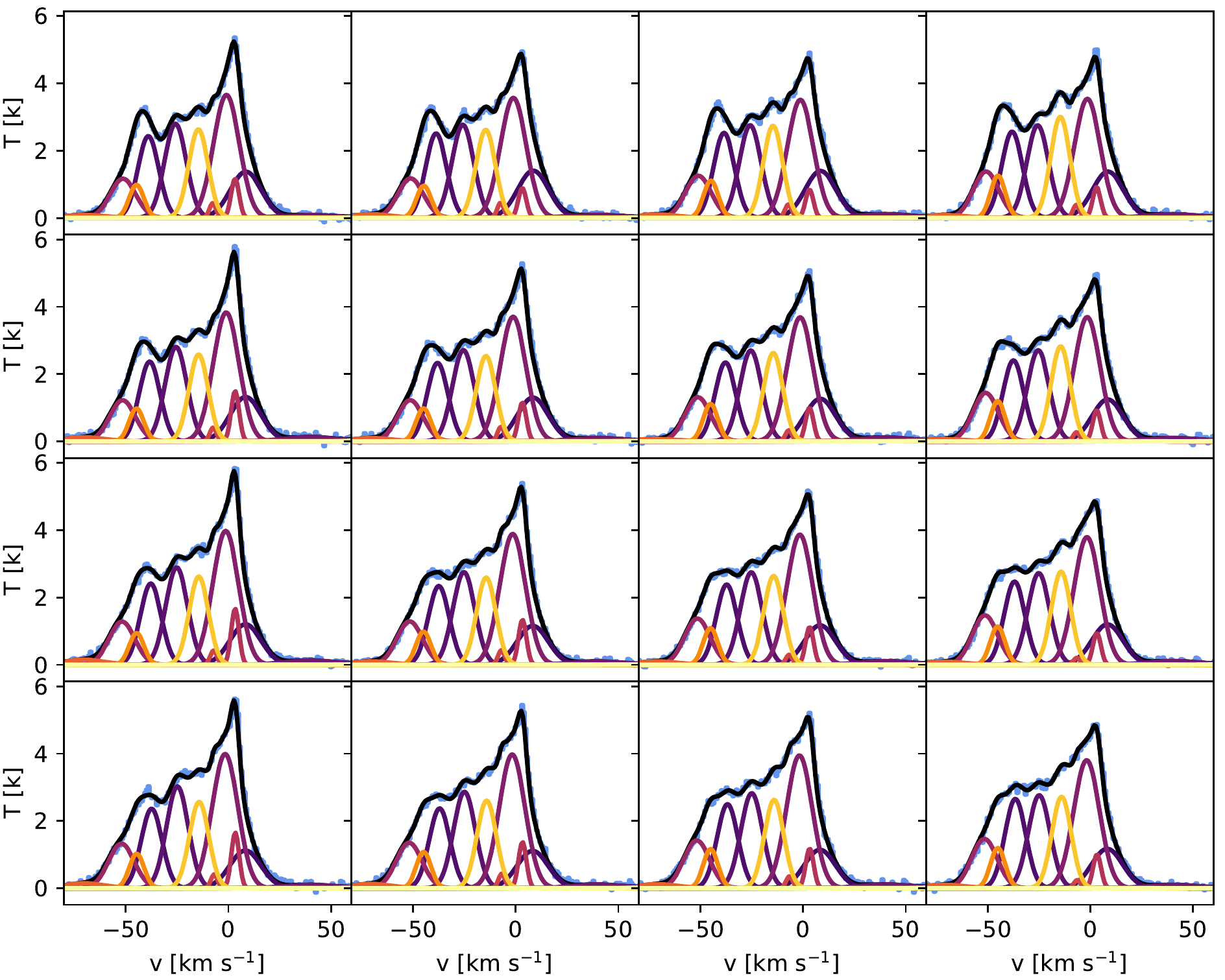}
  \caption{Example of the Gaussian decomposition obtained by {\tt ROHSA}
  (colored line) for a random 4x4 mosaic of NEP. The original signal is show
  by the blue histogram and the total brightness temperature encoded by {\tt ROHSA} is shown
  in black. The other lines detail the components of the Gaussian model. The spatial coherence of the solution can be seen over the mosaic with a smooth variation of the amplitude, central velocity, and velocity dispersion of each Gaussian component.}
  \label{fig::mosaic_spectra_all_NEP}
\end{figure*}

To avoid the complication of low-latitude observations, where the 21\,cm emission is significantly affected by velocity crowding and self-absorption, we chose to apply {\tt RHOSA} on one of the high Galactic latitude fields of the GHIGLS\footnote{\url{http://www.cita.utoronto.ca/GHIGLS/}} survey \citep{martin_ghigls_2015}. We chose the North Ecliptic Pole field (NEP), a 12\textdegree $\times$ 12\textdegree region, centered on $l=96^\circ.40$, $b=30^\circ.03$, observed with the Green Bank Telescope, providing a 9'.55 spatial resolution. The HI spectra have an effective velocity resolution of 0.807 km s$^{-1}$ and cover
-200 < v [km s$^{-1}$] < 50. The integrated column density map computed using Eq.~\eqref{eq::NHI} is shown Fig.~\ref{fig::NHI_TOT_NEP} and a mosaic of representative emission spectra is shown in Fig.~\ref{fig::mosaic_spectra_all_NEP}.

As Fig.~\ref{fig::mosaic_spectra_all_NEP} shows, high-latitude spectra of HI are more complex that the synthetic observations computed from the numerical simulations of \citet{saury_structure_2014}. 
This is caused by a combination of the longer line of sight in the observation (about 200\,pc at $b=30^\circ$ compared to the 40\,pc box of the simulation) and to the presence of non-local velocity components. The 21\,cm emission in NEP indeed exhibits significant emission in the intermediate velocity clouds (IVC) and high velocity cloud (HVC) ranges. In this paper we do not consider the velocity channels with HVC emission. We focus on the phase separation of the local velocity cloud (LVC) and IVC components.

The LVC range, between $-20$\,km\,s$^{-1}$ and $+20$\,km\,s$^{-1}$ shows rather smooth emission profiles, with a narrow peak around $-3$\,km\,s$^{-1}$ on top of a wider feature (see Fig.~\ref{fig::mosaic_spectra_all_NEP}). The latter is rather smooth but when inspected in details it shows faint spectral structures on all scales along the velocity axis. The sensitivity of the GHIGLS data is such that these fluctuations of the emission profiles are not due to noise. In fact, they can be followed from one spectrum to the next quite easily. These fluctuations of the emission spectra, at scales of a few km\,s$^{-1}$ reveal the presence of CNM and LNM features on a range of velocities. 

This field was selected because of its representative 21\,cm emission for Galactic latitudes of $b\sim 30^\circ$. The emission features are not too complex neither too simple. In addition, a first Gaussian decomposition of NEP 21\,cm data was performed by \cite{martin_ghigls_2015} which provides an interesting point of comparison. Unlike {\tt ROHSA}, \cite{martin_ghigls_2015} used a method similar to the one described by \citet{haud_gaussian_2000}, that considers only the term $\|L(v_z, \rb, \thetab(\rb))\|_2^2$ in the parameter optimization. We highlight some qualitative comparisons in the following sections.

\subsection{Results}
\label{subsec::results_NEP}
In order to decompose the 21\,cm emission of the NEP field, we used {\tt ROHSA} with $N = 12$
Gaussian components and each hyper-parameter has been set to 1000. As for the previous case, these values
have been chosen empirically following the same methodology as described in Sect.~\ref{subsubsec::free},
allowing us to converge toward a noise-dominated residual with a minimum number of Gaussian. 
Note that the hyper-parameters values are not the same as for the first application presented 
in Sect. \ref{sec::methodology}. The complexity of the underlying signal structure and its signal-to-noise
ratio are the main causes of these differences. However, a detailed understanding of the behaviour of these
hyper-parameters would require testing different values over a large number of observations. Such
exploration is currently complicated by computations limitations. A GPU version of the code is under
development to overcome this main limitation.

The combination of twelve Gaussian produces a solution that recovers 99 \% of the total emission with spatially coherent components.
The total integrated column density encoded by {\tt ROHSA} and the residual between our model and the data are shown in Fig. \ref{fig::NHI_TOT_NEP} (middle and right panels).

\subsubsection{Global properties of the Gaussian sample}
\begin{table*}[]
\centering
\begin{tabular}{c c c c c c c c c c c c c}
\toprule
 & $G_{1}$ & $G_{2}$ & $G_{3}$ & $G_{4}$ & $G_{5}$ & $G_{6}$
 	& $G_{7}$ & $G_{8}$ & $G_{9}$ & $G_{10}$ & $G_{11}$ & $G_{12}$\\ 
\midrule
$\langle\mub_n\rangle$ [km s$^{-1}$] & -74.1 & -53.9 & -44.7 & -35.0 & -22.9 & -12.6 & -4.8 & -1.3 & 0.2 & 10.9 & 40.8 & 75.9 \\
$\langle\sigmab_n\rangle$ [km s$^{-1}$] & 9.7 & 6.2 & 3.5 & 5.1 & 5.3 & 4.7 & 1.6 & 6.3 & 1.9 & 7.5 & 12.5 & 9.3 \\
\bottomrule
\end{tabular}
\caption{Mean velocity $\langle\mub_n\rangle$ and mean velocity dispersion
	$\langle\sigmab_n\rangle$ of the twelve Gaussian components $G_n$ inferred
    by {\tt ROHSA} in NEP.}
\label{table::mean_var_NEP}
\end{table*}

\begin{figure*}[]
  \centering
  \includegraphics[width=\textwidth,height=\textheight,keepaspectratio]{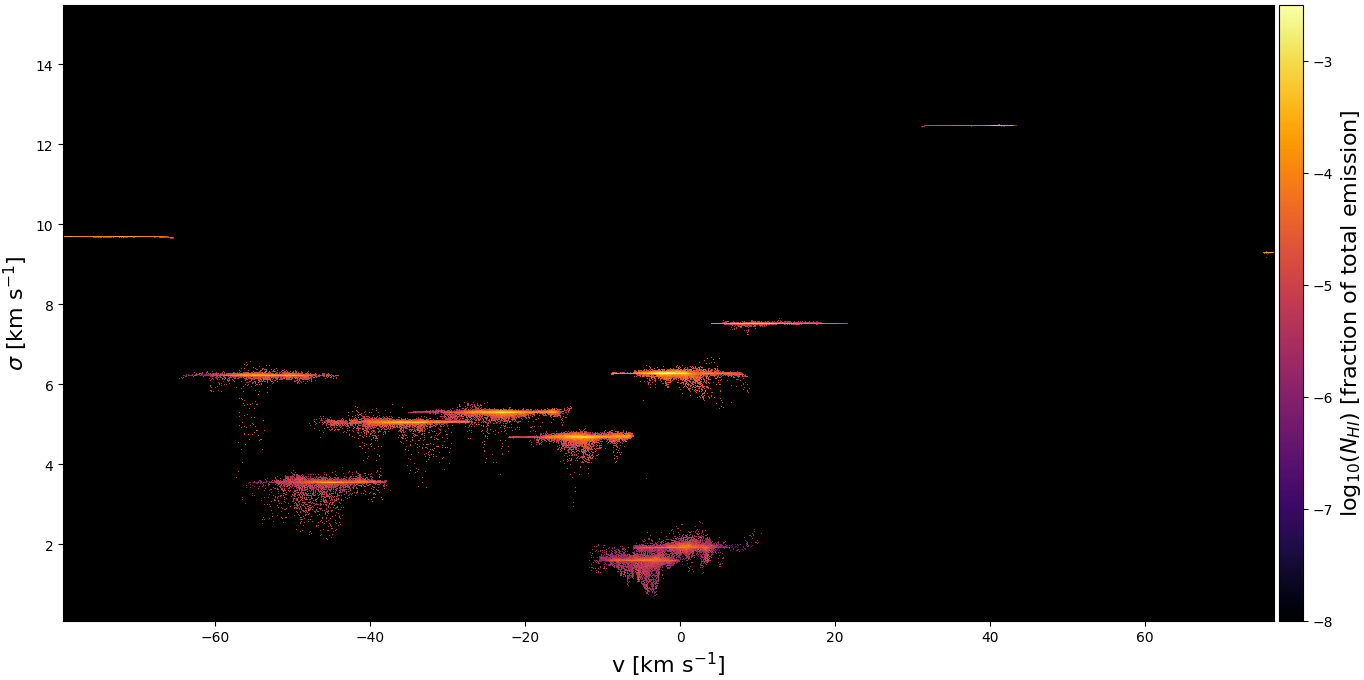}
  \caption{Two dimensional probability distribution function
  $\sigma$-v weighted by the fraction of total emission of each Gaussian 
  $\sqrt{2\pi}\ab_n\sigmab_n/\sum_{\rb}^{}N_{HI}(\rb)$ of NEP.
  NEP is mainly composed of negative intermediate velocity components. A clear
  trend is visible in the velocity range -60 < $\bar v$ [km s$^{-1}$] < 0 with
  velocity dispersions decreasing from 10 km s$^{-1}$ to 1 km s$^{-1}$ going
  from negative to positive velocities. It reflects directly the radiative
  condensation of warm intermediate velocity clouds into the local velocity
  component of the neutral ISM.}
  \label{fig::heatmap_NEP}
\end{figure*}

Like for the application on the synthetic observations, the Gaussian parameters recovered for the NEP field have a strong spatial coherence; {\tt ROHSA} converges towards a solution with smooth variations of the Gaussian parameters across the field. 
It turns out that {\tt ROHSA} converges toward a multiphase model with Gaussian components of various widths, very similar to the application to numerical simulations presented in Section~\ref{subsec::results} but more complex due to the presence of an IVC component in the data.

To have a global view of the thermal state of the gas as a function of velocity, it is useful to look at the two-dimensional dispersion-velocity diagram $\sigmab-v$ weighted by the fraction of total emission of each Gaussian $\sqrt{2\pi}\ab_n\sigmab_n/\sum_{\rb}^{}N_{HI}(\rb)$ shown in Fig. \ref{fig::heatmap_NEP}. 
This diagram  shows isolated complexes of Gaussian components in the $\sigmab-v$ space. This is a direct result of the way the parameter optimization is done in {\tt ROHSA}, with a regularization term that favours the minimum variance of $\sigmab$. 
Figure~\ref{fig::heatmap_NEP} highlights the fact that the 21\,cm emission in NEP is mainly composed of negative velocity components. A clear trend is visible in the velocity range $-60 < v < 0$\,km\,s$^{-1}$ with $\sigmab$ decreasing from 10\,km\,s$^{-1}$ to 1\,km\,s$^{-1}$ going from negative to positive velocities. It likely reflects the radiative condensation of warm intermediate velocity clouds into the local velocity component of the neutral ISM.

At this point it is interesting to compare the results of {\tt ROHSA} with the Gaussian decomposition of the same data performed by \citet{martin_ghigls_2015}. 
The $\sigmab-v$ diagram of \citet{martin_ghigls_2015} (see their Fig.~7) shows a continuous distribution with arches that bridge together LVC and IVC gas, an effect that is not observed in our results.  
Similarly to what we have done here, \citet{martin_ghigls_2015} used numerical simulations to evaluate the performances of their Gaussian decomposition algorithm. Their tests revealed that such arches in the $\sigmab-v$ diagram are unphysical. They are the result of LVC and IVC gas components that overlap in velocity. 

It is important to point out that both solutions are providing as good a representation of the same dataset. The significant difference between the two solutions highlights the challenge of extracting a physically meaningful representation of the data. 
We recall that \citet{martin_ghigls_2015} used an algorithm similar to the ones used by \cite{haud_gaussian_2000,miville-deschenes2017,kalberla_properties_2018} where the spatial coherence of the solution is not enforced through regularization terms in the cost function. In these previous studies spatial coherence is attempted by providing spatially coherent initial guesses. Then each spectrum is fitted independently and no spatial coherence in the solution is enforced. In practice, this method is rather effective for relatively sparse emission data like CO \citep{miville-deschenes2017}, but in the case of the more confused 21\,cm data it produces parameter maps that are more affected by small scale noise due to the degeneracy of the solution.

A Gaussian decomposition algorithm that fits each spectrum individually is easily fooled by components that overlap in velocity. In this specific case, such an algorithm would find a solution with a smaller number of components but with larger values of $\sigmab$. The main innovation in {\tt ROHSA} is that it is able to cluster different phases even if they are close in velocity. The four energy terms added to the cost function $J(\thetab(\rb))$ allow {\tt ROHSA} to find a spatially coherent solution while avoiding the mix of components due to the high confusion present in the emission. 

\subsubsection{A cloud/inter-cloud medium vision of the North ecliptic pole}
\begin{figure*}[]
  \centering
  \includegraphics[width=0.95\textwidth,height=0.95\textheight,keepaspectratio]{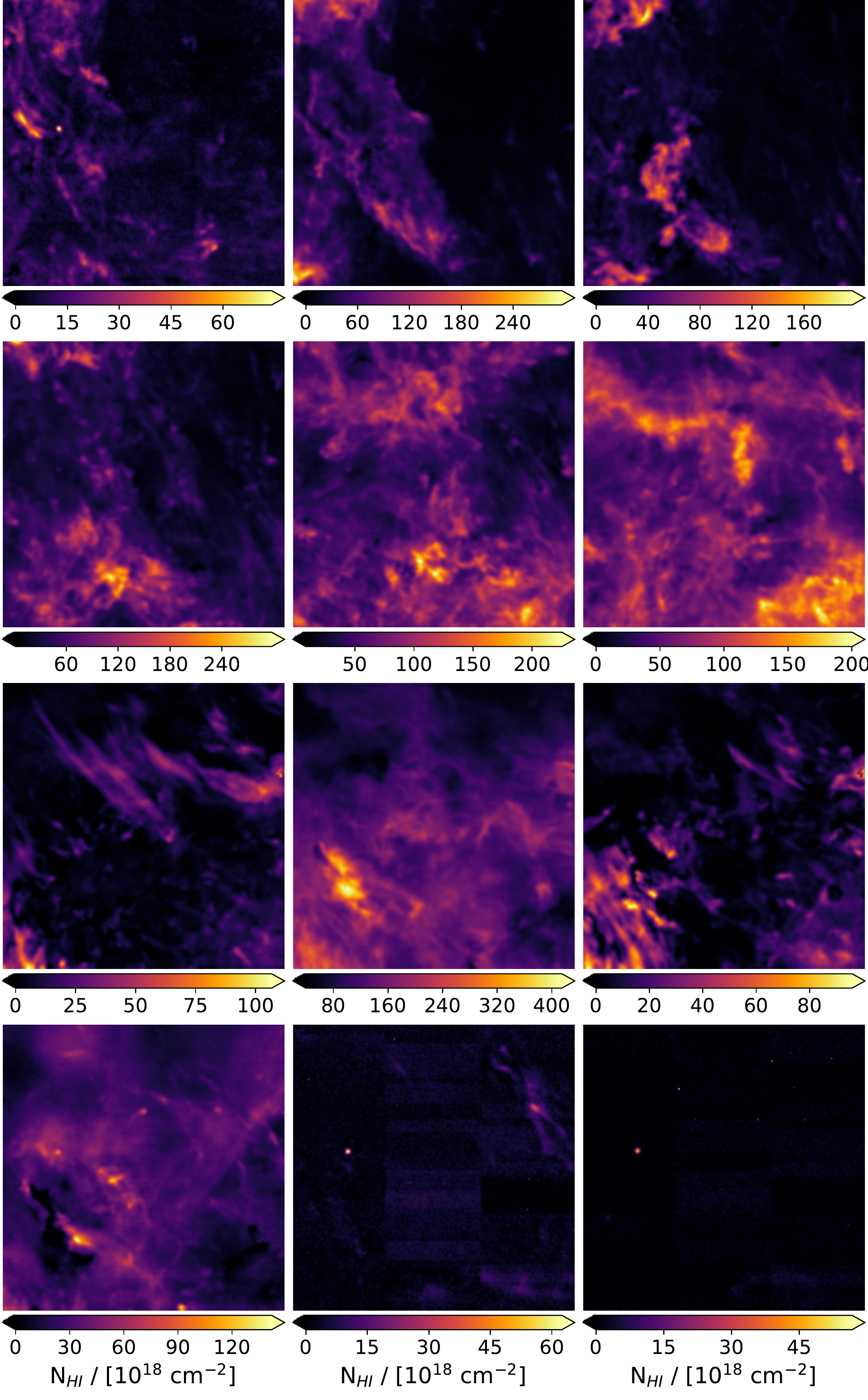}
  \caption{Integrated column density maps (left: $G_1$, $G_4$, $G_7$,
  $G_{10}$; middle: $G_2$, $G_5$, $G_8$, $G_{11}$; right: $G_3$, $G_6$,
  $G_9$, $G_{12}$) obtained by {\tt ROHSA} applied on NEP.
  Mean velocity $\langle\mub_n\rangle$ and mean velocity dispersion
  $\langle\sigmab_n\rangle$ are presented in Table \ref{table::mean_var_NEP}. }
  \label{fig::mosaic_field_NEP}
\end{figure*}

\begin{figure*}[]
  \centering
  \includegraphics[width=0.95\textwidth,height=0.95\textheight,keepaspectratio]{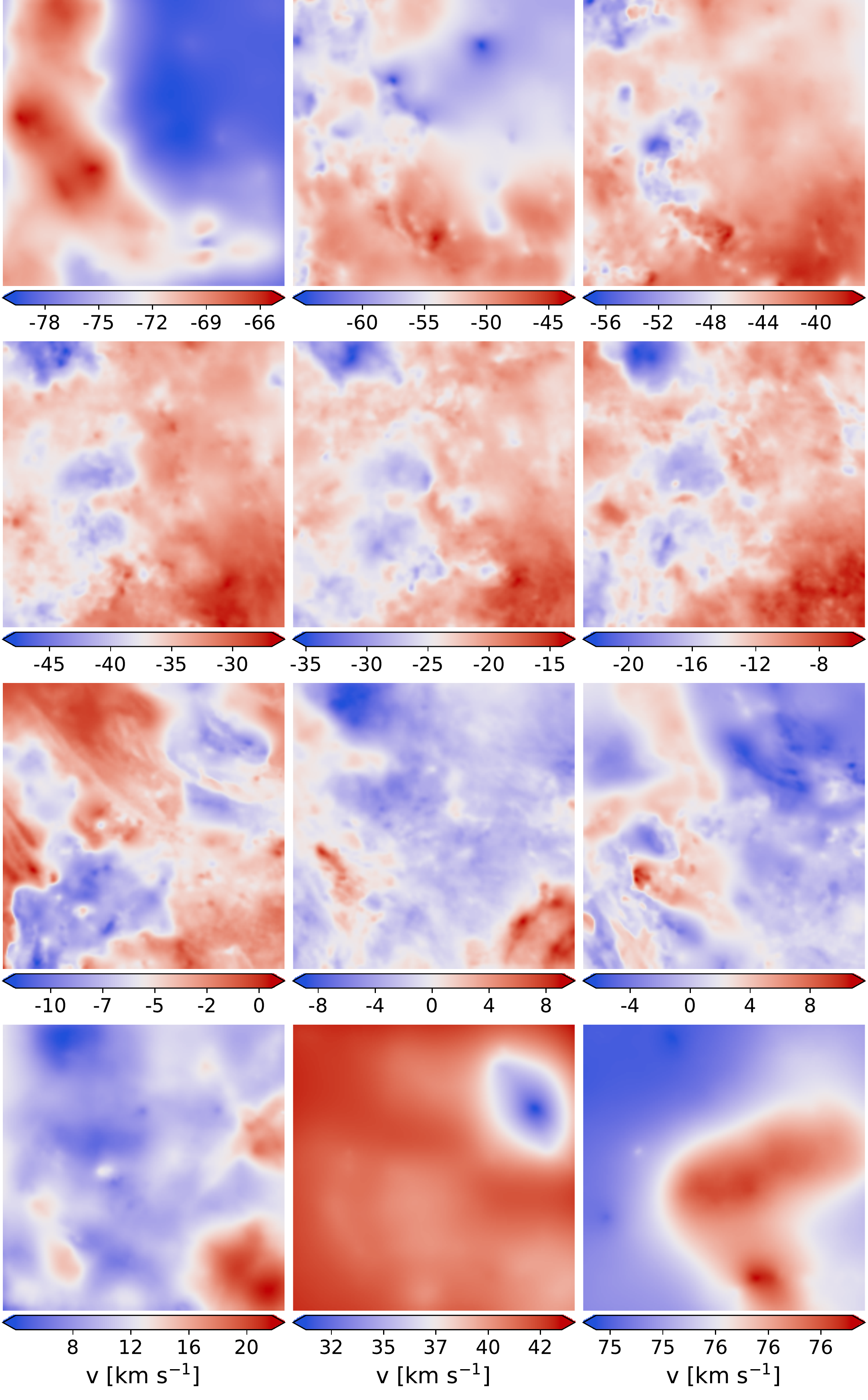}
  \caption{Centroid velocity fields $\mub$ (left: $\mu_1$, $\mu_4$, $\mu_7$,
  $\mu_{10}$; middle: $\mu_2$, $\mu_5$, $\mu_8$, $\mu_{11}$; right: $\mu_3$, $\mu_6$,
  $\mu_9$, $\mu_{12}$) obtained by {\tt ROHSA} applied on NEP.}
  \label{fig::mosaic_vfield_NEP}
\end{figure*}

\begin{figure*}[]
  \centering
  \includegraphics[width=0.95\textwidth,height=0.95\textheight,keepaspectratio]{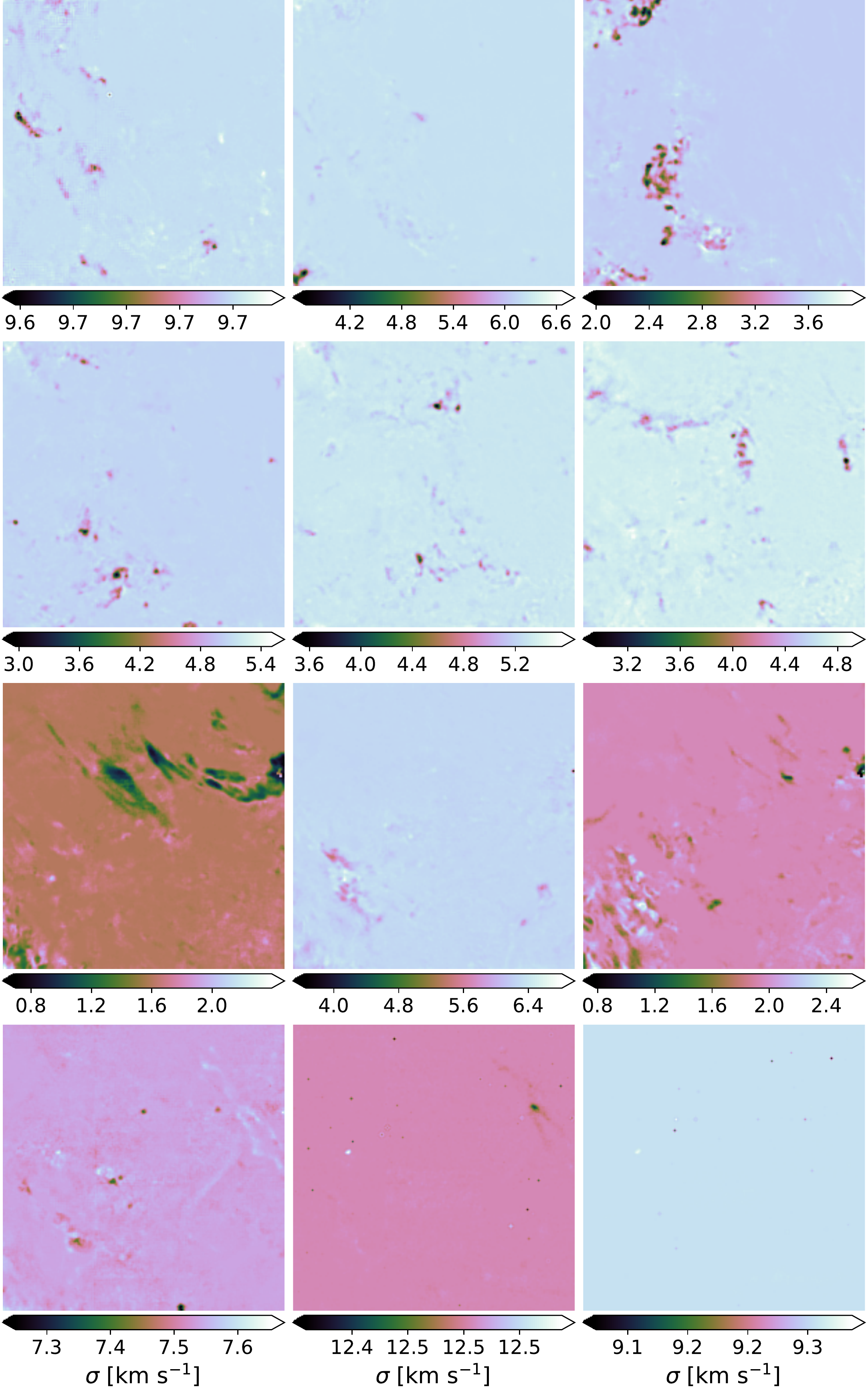}
  \caption{Velocity dispersion maps $\sigmab$ (left: $\sigma_1$, $\sigma_4$, $\sigma_7$,
  $\sigma_{10}$; middle: $\sigma_2$, $\sigma_5$, $\sigma_8$, $\sigma_{11}$; right: $\sigma_3$, $\sigma_6$,
  $\sigma_9$, $\sigma_{12}$) obtained by {\tt ROHSA} applied on NEP.}
  \label{fig::mosaic_sigfield_NEP}
\end{figure*}

\begin{figure}[]
  \centering
  \includegraphics[width=\linewidth]{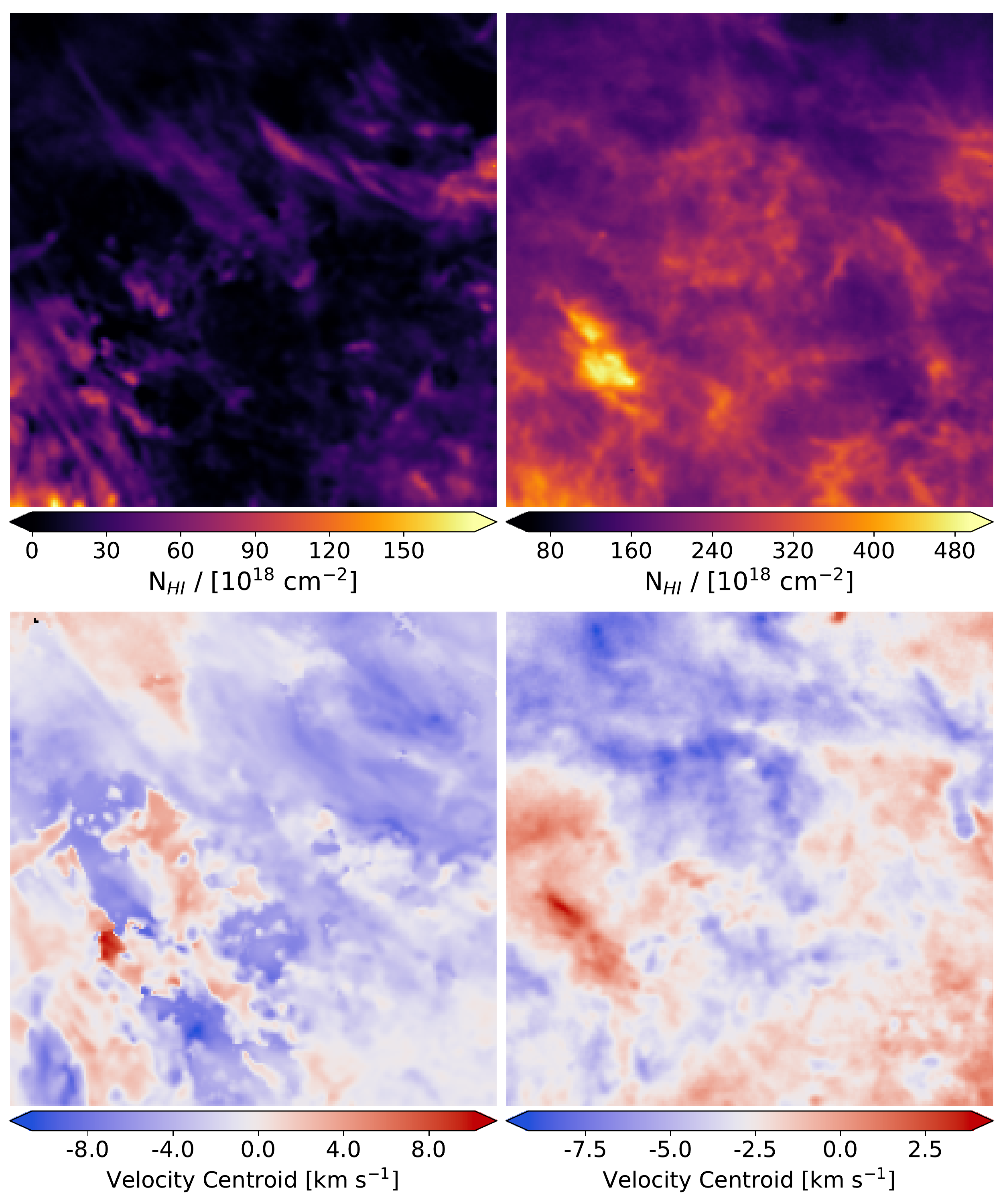}
  \caption{Left : CNM; Right : inter-cloud medium (WNM+LNM) in NEP inferred with {\tt ROHSA}. Top and bottom panels show the column density and centroid velocity fields respectively.} 
  \label{fig::NHI_CV_INTER_NEP}
\end{figure}

Integrated column density maps, centroid velocity fields and dispersion
velocity fields obtained with {\tt ROHSA} are respectively presented in Figs.
\ref{fig::mosaic_field_NEP}, \ref{fig::mosaic_vfield_NEP} and
\ref{fig::mosaic_sigfield_NEP}. Mean velocities $\langle\mub_n\rangle$ and mean
velocity dispersions $\langle\sigmab_n\rangle$ of the twelve Gaussian components G$_n$ are
presented in Table \ref{table::mean_var_NEP}. In this section we focus on building 
a coherent cloud/inter-cloud medium vision considering the local component of the
emission identified previously. 
Two of the four components of the local gas, $G_7$ and $G_9$,
are associated with the CNM with $\langle\sigmab_7\rangle$ = 1.6 km
s$^{-1}$ and $\langle\sigmab_9\rangle$ = 1.9 km s$^{-1}$. The other ones are used to build the inter-cloud medium. Integrated column density fields and centroid velocity fields of the cloud medium and inter-cloud medium are presented Fig. \ref{fig::NHI_CV_INTER_NEP}. 

As noted by \cite{martin_ghigls_2015} in their two-phase
decomposition of the local component, filamentary structures are observed in the narrow component. The associated velocity dispersion fields (see Fig.~\ref{fig::mosaic_sigfield_NEP}, component $G_7$ and $G_9$) shows coherent fluctuating values over a large part of the field. The core of these filamentary structures appears narrower than the envelop with velocity dispersion reaching about 0.87 km s$^{-1}$ (the spectral resolution) in their centers.  

The wider component has an integrated column density field with
no particular structure like filaments (see Fig.~\ref{fig::NHI_CV_INTER_NEP}, top-right).  Like for the numerical simulation, the sum of the wide components is likely to represent a phase that fills a large fraction of the volume, as would an inter-cloud medium. 
One interesting aspect of the {\tt ROHSA} decomposition is that it then allows to extract the velocity field of this volume filling component (Fig.~\ref{fig::NHI_CV_INTER_NEP}, bottom-right), enabling the characterization of the turbulent cascade in a mixture of lukewarm phase and warm phase.  

\section{Discussion}
\label{sec::discussion}

Historically a large number of studies used a Gaussian basis to model 21\,cm data. Different algorithms have been developed, all of them are fitting each spectrum individually, with or without information from the neighbouring solutions to initialize the fit. To constrain further the degeneracy of the fit, solutions with the smallest number of Gaussian components have often been favoured \citep[e.g.,][]{lindner_autonomous_2015}. Because of velocity blending, the solution with the smallest number of components is not necessarily the best one. In some cases, narrow features overlap in velocity, making it impossible to separate them if the environment is not considered. Usually, this confusion breaks apart a few beams away and more components can be recovered. The fundamental idea behind {\tt ROHSA} is that we are trying to extract diffuse components that have column density, centroid velocity and velocity dispersion with smooth spatial variations. The optimization scheme has been designed with that concept at its core. In order to achieve this, {\tt ROHSA} fits the whole data cube at once.

The application of {\tt ROHSA} on both synthetic observations from numerical simulations and observational data converges naturally toward a multiphase model of the neutral ISM. The ability of {\tt ROHSA} to extract the multiphase nature of the neutral ISM opens a totally new perspective on the study of the nature of the condensation process acting in the ISM. It is clear from the numerical point of view that the formation of cold clouds is the result of the condensation of the warm and diffuse gas through the thermal instability coupled with turbulence. From the observational point of view, spatial correlations between the different phases can now be made in order to quantify more precisely how the CNM emerges from this condensation process. This separation also opens the possibility to describe the properties of the very specific multiphase turbulence of the HI.

{\tt ROHSA} appears to be efficient at clustering different structures in PPV space, even when there is a high level of confusion. In particular, the separation of the LVC and IVC is known to be particularly challenging as the CNM and WNM of both components overlap significantly in velocity \citep{martin_ghigls_2015}. As shown in Fig. \ref{fig::heatmap_NEP}, {\tt ROHSA} limits significantly the "Arch effect" typical of this confusion (see Sect.~\ref{sec::application}). 
Globally the performance in {\tt ROHSA} on high Galactic latitude regions opens a large range of possibilities regarding the study of infalling neutral clouds from the galactic halo.

We would also like to point out that no a-priori informations about the number of phases present in the neutral ISM is provided to {\tt ROHSA}. The algorithm rests only on the hypothesis of the existence of components with similar line-width through the energy term $\lambda'_{\sigmab} \|\sigmab_n - m_n\|_2^2$. In that respect, {\tt ROHSA} is perfectly adapted to decompose hyper-spectral observations of any type, not only 21\,cm emission.

At this time the main limitations of {\tt ROHSA} are computational. Firstly, as the whole PPV cube is fitted at once, the use of {\tt ROHSA} is limited to cubes that can fit in memory. Secondly, the current computation time of {\tt ROHSA} is not negligible (e.g., about two hours on a single CPU for a $256\times256\times 100$ PPV cube with 8 Gaussians). This limits the possibility to make a deep exploration of the hyper-parameters $\lambda_i$. In particular it would be interesting to explore various balance of the hyper-parameters for the deduced quantities ($a$, $\mu$ and $\sigma$). It is expected that the amplitude of the spatial variations of these quantities are not the same. For instance, in a multi-phase medium like the HI the density field (represented by $a$) might vary more strongly at small scale than the velocity field. This might require different values of $\lambda_a$ compared to the others. A GPU verion of the code is under development that would allow such an exploration.

\section{Summary}
\label{sec::summary}
In this study we have presented a new Gaussian decomposition algorithm named
{\tt ROHSA}. Energy terms have been added to the classical
cost function to take into account simultaneously the spatial coherence of
the emission and the multiphase nature of the gas. In order to identify a solution with spatially smooth parameters, the fit is performed on the whole hyper-spectral cube at once. 

The performance of {\tt ROHSA} has been evaluated using a synthetic 21\,cm observation
computed from a numerical simulation of thermally bi-stable turbulence. It was then tested on a 21\,cm observation of a high Galactic latitude field observed with the GBT. 
The main conclusions are the following :

\begin{enumerate}
	\item {\tt ROHSA} is able to highlight naturally the physics of
    any multiphase medium without a-priori regarding the number of phase.

    \item Evaluation on numerical simulation of thermally bi-stable turbulence
    shows that the sum of Gaussian components is a good approximation to model
    the multiphase nature of the neutral ISM.

    \item The multiphase model inferred with {\tt ROHSA} provides a spatially
    coherent vision of the integrated column density map, the centroid velocity
    field and the velocity dispersion field of each component.

    \item The power spectra of the integrated column density
    and centroid velocity fields are well recovered with {\tt ROHSA}.
    Statistical properties of turbulence in the multiphase neutral ISM now become
    accessible.

    \item The decomposition of a high-latitude HI gas observation shows the wide range
    of applications enabled with {\tt ROHSA}, for instance to study the radiative
    condensation of the WNM and the nature of the ISM at the disk-halo interface.
\end{enumerate}

\appendix
\section{Optimization algorithm}
\label{app:gradient}
Terms used to compute the gradient $\nabla J(\thetab, \mb)$ of the cost function $J(\thetab, \mb)$ are detailed in this appendix. It involves the Jacobian of the residual  $\nabla L(\thetab)$ and the gradients of the regularization term 
$\nabla R^t(\thetab, \mb) = [\nabla_{\thetab} R^t(\thetab,\mb), \nabla_{\mb} R^t(\thetab,\mb)]$

\begin{ceqn}
\begin{align}
    \nabla L(v_z, \thetab(\rb)) = 
    \begin{bmatrix}
        \exp \left( - \frac{\big(v_z - \mu_1(\rb)\big)^2}{2 \sigma_1(\rb)^2} \right) \\
        \frac{a_1 (v_z - \mu_1(\rb))}{\sigma_1^2} \exp \left( - \frac{\big(v_z - \mu_1(\rb)\big)^2}{2 \sigma_1(\rb)^2} \right) \\
        \frac{a_1 (v_z - \mu_1(\rb))^2}{\sigma_1^3} \exp \left( - \frac{\big(v_z - \mu_1(\rb)\big)^2}{2 \sigma_1(\rb)^2} \right) \\
        \vdots \\
        \frac{a_N (v_z - \mu_N(\rb))^2}{\sigma_N^3} \exp \left( - \frac{\big(v_z - \mu_N(\rb)\big)^2}{2 \sigma_1(\rb)^2} \right)
    \end{bmatrix}
\end{align}
\end{ceqn}

\begin{ceqn}
\begin{align}
  \label{eq:7}
  \nabla_{\thetab} R(\thetab, \mb) = 
  \begin{bmatrix}
    \lambda_{\ab} \Db^t \Db  \ab_1  \\
    \lambda_{\mu} \Db^t \Db  \mub_1 \\
    \lambda_{\sigmab} \Db^t \Db \sigmab_1 \\
    \lambda'_{\sigmab} (\sigmab_1 - m_1) \\
    \vdots \\
    \lambda'_{\sigmab} (\sigmab_N - m_N) \\
  \end{bmatrix}
\end{align}
\end{ceqn}

\begin{ceqn}
\begin{equation}
    \nabla_{\mb} R(\thetab, \mb) = 
    \begin{bmatrix}
        - \sum_{\rb} \lambda'_{\sigmab} (\sigmab_1 - m_1) \\
        \vdots \\
        - \sum_{\rb} \lambda'_{\sigmab} (\sigmab_N - m_N) 
    \end{bmatrix}
\end{equation}
\end{ceqn}

\section{Computation time}
\label{app::timestep}
\begin{figure}[]
  \centering
  \includegraphics[width=0.95\linewidth]{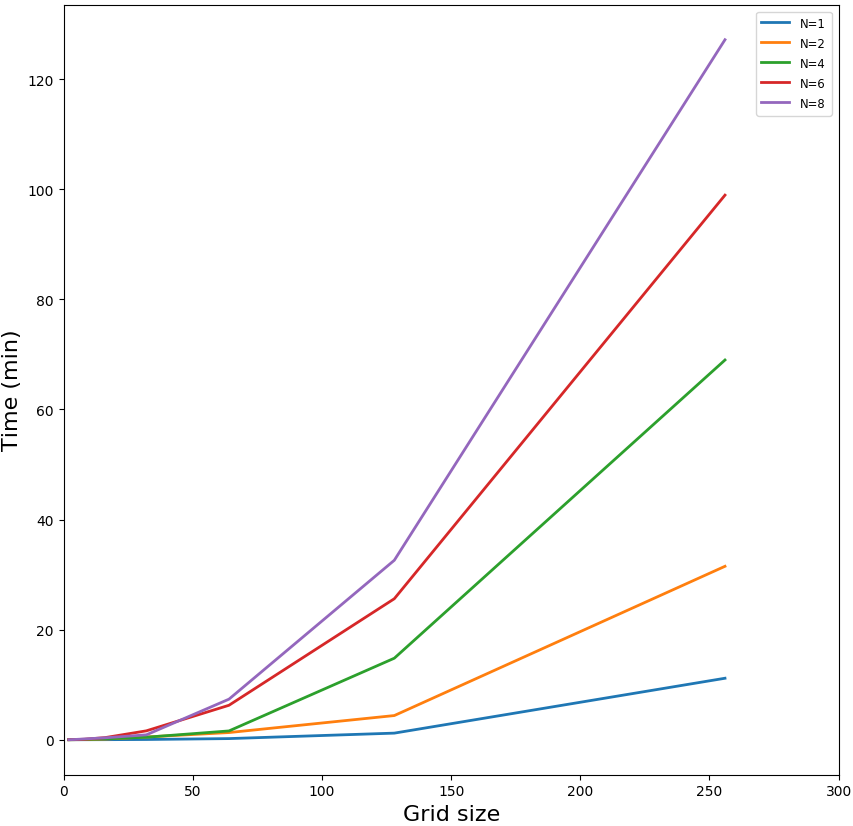}
  \caption{Computation time used by one CPU to perform the Gaussian decomposition of the simulated PPV cube used in Sect. \ref{sec::evaluation}, for N=1,2,4,6,8 Gaussian, as function of the size grid. The maximum number of iteration in each case has been set to 800.}
  \label{fig::timestep}
\end{figure}

The computation time needed to perform the Gaussian decomposition of the synthetic PPV cube presented in Sect. \ref{sec::evaluation} is described in Fig.~\ref{fig::timestep}. The computation time depends on the number of Gaussian component $N$, on the size of the cube (number of spectra and number of velocity channels) and on the maximum number of iteration used in the optimization.

For a given number of component $N$, the computation time scales linearly with the number of spectra and number of velocity channels. Therefore, as for each step of the multi-resolution process, from coarse to fine grid, the size of the grid is multiplied by a factor of four, the computation time also increases by a factor four at each step.  
Note that each step has the same maximum number of iteration (here 800) that impact also linearly the computation time.

Finally, we observed that the computation time depends non-linearly on the number of Gaussian $N$, as seen in Fig.~\ref{fig::timestep}.

\begin{acknowledgements}
	Part of this work was supported by Hyperstars, a project funded by the MASTODONS initiative of the CNRS mission for inter-disciplinarity and by the Programme National “Physique et Chimie du Milieu Interstellaire” (PCMI) of CNRS/INSU with INC/INP co-funded by CEA and CNES. 
	This work took part under the program Milky-Way-Gaia of the PSI2 project funded by the IDEX Paris-Saclay, ANR-11-IDEX-0003-02.
    We gratefully acknowledge Lucie Riu for enlightening conversations.
    We thank the anonymous referee whose comments and suggestions have improved this manuscript.
\end{acknowledgements}
\bibliographystyle{aa}
\bibliography{ROHSA}

\appendix
\end{document}